\documentclass{aa} 
\usepackage{graphicx}
\usepackage{txfonts}
\usepackage{natbib}
\usepackage{color}
\usepackage{array}
\usepackage{subfig}
\usepackage{xspace}
\usepackage{url}

\newcommand{\equ}[1]{Eq.~\ref{eq:#1}}
\newcommand{\fig}[1]{Fig.~\ref{fig:#1}}
\newcommand{\tab}[1]{Tab.~\ref{tab:#1}}
\newcommand{\sect}[1]{Sect.~\ref{sec:#1}}

\newcommand{\lcdm}[0]{$\Lambda$CDM\xspace}

\begin{document}

\title{MOND simulation suggests the origin of some peculiarities in the Local Group}

\author{M. B\'{i}lek\inst{1}\fnmsep\inst{2}\fnmsep\inst{3}
\and
I. Thies\inst{2}
\and
P. Kroupa\inst{2}\fnmsep\inst{3}
\and
B. Famaey\inst{4}
}
\institute{Astronomical Institute, Czech Academy of Sciences, Bo\v{c}n\'{i} II 1401/1a, CZ-141\,00 Prague, Czech Republic\\
\email{bilek@asu.cas.cz}
\and
Helmholtz-Institut f\" ur Strahlen und Kernphysik, Nussallee 14-16, 53115 Bonn, Germany
\and 
Charles University in Prague, Faculty of Mathematics and Physics, Astronomical Institute, V Hole\v sovi\v ck\' ach 2, 180 00 Prague 8, Czech Republic
\and
 Universit\'e de Strasbourg, CNRS UMR 7550, Observatoire astronomique de Strasbourg, 11 rue de l’Universit\'e, 67000 Strasbourg, France
}

\date{Received ...; accepted ...}

\abstract
{The Milky Way (MW) and Andromeda (M\,31) galaxies possess rotating planes of satellites. Their formation has not been explained satisfactorily yet. It was suggested that the MW and M\,31 satellites are ancient tidal dwarf galaxies, which could explain their configuration. This {suggestion} gained support by an analytic backward calculation of the relative MW-M\,31 orbit in the MOND modified dynamics paradigm by Zhao et al. (2013) implying their close flyby 7-{11}\,Gyr ago.
}
{{Here we explore the Local Group history in MOND in more detail using a~simplified first-ever self-consistent simulation. We note the features induced by the encounter in the simulation and identify their possible real counterparts. }}
{ {The initial conditions were set to eventually roughly reproduce} the observed MW and M\,31 masses, effective radii, separation, relative velocity and disk inclinations. We used the publicly available adaptive-mesh-refinement code Phantom of RAMSES.}
{{Matter was transferred from the MW to M\,31 along a~tidal tail in the simulation. The encounter induced formation of several structures resembling the peculiarities of the Local Group. Most notably: 1) A~rotating planar structure formed around M\,31 from the transferred material. It had a~size similar to the observed satellite plane and was oriented edge-on to the simulated MW, just as the real one. 2) The same structure also resembled the tidal features observed around M\,31 by its size and morphology. 3)  A~warp in the MW developed with an amplitude and orientation similar to that observed.{ 4) A~cloud of particles formed }around the simulated MW, with the extent of the actual MW satellite system.  The encounter did not end by merging in a~Hubble time. {The simulated stellar disks also thickened as a~result of the encounter.}}}
{The simulation demonstrated that  MOND can possibly explain many peculiarities of the Local Group{, which should be verified by future more elaborate simulations.} The simulation moreover showed that tidal features observed in galaxies, usually interpreted as merger remnants, could have been formed by matter exchange during non-merging galactic flybys in some cases. }

\keywords{
Galaxies: Local Group --
Galaxies: interactions --
Galaxies: formation --
Galaxies: kinematics and dynamics --
Galaxies: structure --
Gravitation
}

\maketitle

\section{Introduction} \label{sec:intro}
The satellites of the Milky Way (MW) have a~remarkable spatial distribution \citep{lynden-bell76, kroupa05, metz07}: Their positions define a~flattened body (a satellite plane, SP, called the Vast Polar Structure, VPOS, {in the case of the MW}) with a~root-mean-square (RMS) half-thickness of around 15\,kpc and a~RMS radius of around 40\,kpc \citep{pawlowski15newsat}. The most distant satellite lies 365\,kpc away from the MW center. The central plane of this cloud is almost perpendicular to the MW disk and almost goes through the MW center \citep{kroupa05}. The angle between the VPOS and the line connecting the centers of the MW and the Andromeda galaxy (M\,31) is around 40-50\degr \citep{pawlowski13}. The velocities of the satellites are mostly consistent with orbiting within the SP \citep{metz08}. \citet{pawlowski13vpos} found that for the 11 brightest MW satellites, 9 orbit within the SP. Of them, 8 orbit the MW in the same sense and 1 in the opposite. Not only the satellites, but also stellar streams and outer halo globular clusters are concentrated within the VPOS \citep{pawlowski12}. The classical, bright, satellites are more concentrated toward the midplane of the VPOS than the ultra-faint dwarfs \citep{kroupa12}.

These discoveries motivated the search for {SPs} in other galaxies. A~spectacular example was found in M\,31, called the Great Plane of Andromeda, GPoA, \citep{metz07, metz09, ibata13}. \citet{ibata13} revealed that 15 out of the 27 satellites with distances known at that time formed a~plane pointing to the MW by its edge. This orientation enabled determining, without the knowledge of the tangential velocities, that 13 of the 15 satellites in the plane were consistent with co-rotating around M\,31. This GPoA rotates in the same sense as the VPOS. \citet{pawlowski13} used a~newer data set to conclude that up to 19 out of 34 M\,31 satellites contribute to the plane. From the values given by \citet{pawlowski13}, we could calculate that the inclination of this plane with respect to the line joining M\,31 with the MW is just 2\degr. This SP has a~RMS half-thickness of 14\,kpc and a~RMS radius of around 130\,kpc along its long axis and 25\,kpc along its intermediate axis \citep{pawlowski13}. It is nearly perpendicular to the MW galactic disk. Most of the M\,31 satellites belonging to the GPoA lie on the side of M\,31 closer to the MW, i.e., their distribution is lopsided. This remarkable property of the M\,31 satellites is not exceptional: \citet{libeskind16} stacked images of many observed galaxy pairs and their satellites. They found a~significant tendency for the satellites to lie between the pair.

There are also non-satellite dwarfs in the Local Group (LG) whose distances to the MW and M\,31 are comparable. \citet{pawlowski13} looked for planes in the whole LG and found that 14 of the 15 non-satellite dwarfs lie on two very large and thin planes (RMS half-thickness of around 60\,kpc, RMS half-length of around 600\,kpc along the longest axis). They are both parallel with the line connecting MW and M\,31 and they have equal distances to it, around 150\,kpc. The galactic disk of M\,31 lies on the symmetry plane between the non-satellite dwarf planes. The LG thus has a~high unexplained symmetry.

Some of the LG dwarfs analyzed by \citet{pawlowski13} lie in the disk planes of M\,31 or the MW. 

The nearest big galaxy outside the LG is Centaurus~A (Cen~A, NGC\,5128). \citet{tully15} reported a~discovery of two {almost parallel} SPs here seen edge-on from the MW, both with the RMS half-thickness of 60\,kpc and a~RMS radius of 350\,kpc. \citet{muller16} made another analysis of this satellite system taking into account also newly discovered faint satellites. They concluded that the hypothesis that there is only a~single thick SP could not be excluded. {\citet{muller18} has recently confirmed that most satellites in this plane co-rotate.} A~tidal stream extends from one of the Cen~A satellites. The distance to several spots in the stream was measured by \citet{crnojevic16}. \citet{muller16} revealed that the stream lies almost in one of the original SPs by \citet{tully15}.

The next nearest galaxy group is the M\,81 group. Here \citet{chiboucas13} noted that the early-type dwarf galaxies are distributed in a~flattened formation tilted toward the line of sight, while the distribution of the late-type dwarfs is more isotropic.

The last SP known to us is that around M\,101 reported by \citet{muller17} recently. It is again seen edge-on from the MW.

\citet{ibata14} estimated statistically the occurrence of rotating SPs in more distant galaxies in the SDSS database. A~galaxy was counted as a~candidate for having a~rotating SP if the positions and radial velocities of its satellites were consistent with such a~hypothesis. They found that the frequency of these candidates is greater than what one would expect if the galaxies had their satellites distributed isotropically. Their result is consistent with a~fraction of about 50\% of
satellites residing in SPs around M31 and MW-like hosts at redshift z<0.05. \citet{ibata15} made a~similar analysis: They counted a~galaxy as a~candidate for having a~rotating SP, if for some of its spectroscopically confirmed satellite there was a~satellite lying on the opposite side of the galaxy. Again, a~$3\sigma$ overabundance of such candidates was detected.

The present attempts to explain the existence of SPs in the cosmological \lcdm model were not satisfactory (e.g., \citealp{pawlowski15persist}{, \citealp{pawlowski17}}). Cosmological \lcdm dark-matter-only simulations result in {a nearly} isotropic satellite distribution around the host galaxy with only a~mild degree of ellipticity and no preferred orbital direction. \citet{kroupa05} calculated that the probability that the positions of the 11 then known MW satellites are consistent with cosmological dark-matter-only simulations is 0.5\%. {Using the most updated data, \citet{pawlowski16} found that the spatial structure is a~$\approx 5 \sigma$ event with respect to an isotropic distribution. Including the clustering of the orbital momentum vectors of the satellites increases the significance of the VPOS even more.} The probability that a~SP like the GPoA is found among satellites with an isotropic distribution of positions and velocities is 0.002\% \citep{ibata13}. It was suggested that satellites accreted along a~cosmological dark matter filament or accreted in a~compact group would form SPs, but these effects are included naturally in cosmological simulations. {\citet{metz09b} showed that the infall of groups of satellites cannot produce the VPOS.} {\citet{pawlowski15persist} concluded that including baryons into cosmological simulation does not help to remove the problem.} Dwarf galaxies in these simulations reside predominantly in  primordial dark matter halos.

There is also another type of dwarf galaxies which do not possess appreciable dark matter halos according to \lcdm. They are tidal dwarf galaxies (TDGs), gravitationally bound objects formed in tidal tails of interacting galaxies {(see, e.g., \citealp{bournaud10, ploeckinger18})}. They were observed many times. Simulations showed two mechanisms of their formation: by the Jeans instability of the gas in tidal tail, or by accumulating a~great amount of gas at the tip of a~tidal tail. Since the material in the tidal tail was stretched from an originally small volume in a~dynamically cold disk of the parent galaxy, the TDGs have to occupy a~small volume in the phase-space according to Liouville's theorem and to form a~phase-space-correlated structures. This is exactly what is observed in the rotating SPs. Simulations showed that TDGs are free of dark matter \citep{barnes92} which seems to contradict the high dynamical mass-to-light ratios of the LG dwarfs. However, the simulations by \citet{kroupa97} demonstrated that dark-matter-free satellites in Newtonian gravity could appear as dark matter rich if they were not in dynamical equilibrium {(see also \citealp{casas12})}. 

\citet{hammer13} presented a~simulation where a~galaxy accreted by M\,31 produced TDGs that formed a~SP around M\,31. {\citet{fouquet12} found using hydrodynamical simulations that if the Giant Stream at M\,31 is to be reproduced by an accreted satellite then collision formed a~tidal tail pointing towards the MW, which could form the VPOS.}  This scenario accounts for many properties of the LG galaxies. To account for the observed high dynamical masses of the satellites, they relied on the argument by \citet{kroupa97}. 

Another possibility to explain the high dynamical masses of the LG dwarfs if they are TDGs is to assume MOND \citep{milg83a, milgcjp}, a~paradigm suggesting that the {commonly-used} laws of gravitational dynamics (at least) have to be modified for low accelerations rather than to extend the well-proven standard model of particle physics by new particles of dark matter. Galaxy observations usually agree well with MOND in various situations (see, e.g., \citealp{famaey12} for a~review). For example, MOND explains successfully the internal dynamics of most regular early and late type galaxies \citep{begeman91, sanders96, deblok98, milg03, milg07, tiret07b, richtler11, gentile11, angus12, milg12, samur14, mcgaugh16, dabringhausen16, lelli17}, the rotation curves of polar rings \citep{lughausen13}, the properties of the Sagittarius stream \citep{thomas17}, or the galaxy-galaxy weak gravitational lensing \citep{milgrom13}. 

\citet{zhao13} (Z13 hereafter) used the MOND two-body-force formula to calculate the history of the MW-M\,31 relative orbit. They used the best estimates on the current baryonic masses, separation, and radial velocity. Despite some recent debates on the exact
value of the tangential velocity \citep{salomon16}, the value they adopted from \citet{vandermarel12} is still the best direct one based on the Hubble Space Telescope. With this small tangential velocity, they found that the MW and M31 had to  have had a~close encounter 7-11 Gyr ago. This opens the possibility that TDGs formed and are observed as the LG dwarfs now. When integrated backwards, the orbits of the Large and Small Magellanic Clouds were close to pericenter almost when the MW and M\,31 were at the pericenter. Dynamical friction during encounters of comparable galaxies is known to be weaker in MOND than in \lcdm from the simulations by \citet{tiret07, nipoti08, combes16} (see also, \citealp{renaud16} {or \citealp{vakili17}) because of the absence of the large and massive dark matter halos \citep{kroupacjp}}. A~close MW-M\,31 encounter in MOND would therefore be more likely to avoid ending in a~merger than in \lcdm. Simulated galaxy encounters in MOND also produce TDGs more easily than their \lcdm counterparts \citep{tiret07, renaud16}. By employing MOND, the equivalent Newtonian dynamical masses of most {MW and M\,31} satellites increase so that they match the observed values \citep{angus08dwarf, serra10, anddwarfi, anddwarfii}. The remaining discrepancies for the least massive objects might be explained if these objects are not in dynamical equilibrium (e.g., \citealp{mcgaugh10, dabringhausen16}){, as suggested by \citet{kroupa97} but using Newtonian simulations.} The rotation curves of the MW \citep{famaey05, mcgaugh08, iocco15}, M\,31 and M\,33 \citep{famaey12} are also reproduced well. Supposing that some TDGs formed during the MW-M\,31 encounter and remained bound to one of the big galaxies, they would form at least a~temporal \citep{fernando17} {SP in the case that the orbital planes of individual satellites around their hosts happen to nearly coincide.} The non-satellite LG dwarfs with their planar distribution could be ejected TDGs. Further observational evidence for this scenario is given in \sect{disc}. We can possibly see a~formation of a~SP {in progress} in the interacting disk galaxy pair ARP\,87. Here one of the tidal tails seems to be wrapped around the other galaxy to form a~disk-like structure\footnote{An image of ARP\,87 is available at this url: \url{https://apod.nasa.gov/apod/ap151209.html}.}.

{These facts had inspired our present work. In this paper, we made a~step beyond the qualitative considerations and analytic calculations and explored the history of the LG in MOND using a~first-ever self-consistent simulation. We noted the features induced by the encounter and compared the simulation to the observations. The simulation was set so that it approximately reproduced the observed MW-M\,31 masses, disk radii, separation, relative radial and tangential velocity and disk inclinations. The approximately correct rotational curves were ensured by using MOND. The simulation showed a~close encounter of the MW and M\,31. The separation between the galaxies was 24\,kpc at the pericenter occurring around 7\,Gyr ago. Mass was transferred from the MW to M\,31. The encounter induced features around the simulated MW and M\,31 similar to the observed tidal features by their morphology and spatial extent. A~rotating planar structure resembling the GPoA formed at the simulated M\,31 which could be seen edge-on from the position of the Sun in the simulation. The encounter induced a~warp in the simulated MW disk similar to that observed and increased the thicknesses of the MW and M\,31 galactic disks. Future more elaborate simulations should verify whether the features formed by a~MW-M\,31 encounter in MOND can be made closely matching the observations.}

This paper is organized as follows: In \sect{desc} we describe our computational methods and the choice of the input parameters. The outcome of the simulation is described in \sect{res}. We compare the outcome with observations and list some limitations of our {approach} in \sect{disc}. We summarize the paper in \sect{sum}.

%
%
%
%
%
%

\section{Description of the simulation}\label{sec:desc}

\subsection{The equations being solved}
According to the modern formulation \citep{milg09}, non-relativistic MOND is any theory obeying these tenets: 1) It contains a~constant with the dimension of acceleration $a_0$, 2) In the limit $a_0\rightarrow 0$ (all quantities with the dimension of acceleration are much greater than $a_0$) the equations of the theory reduce to Newtonian dynamics, and 3) In the limit of $a_0\rightarrow \infty$ and $G\rightarrow 0$ (all quantities with the dimension of acceleration are much smaller than $a_0$), keeping the product $a_0G$ constant, the dynamics of purely gravitationally interacting systems is space-time scaling invariant, i.e. if the equations of the theory allow the bodies to move on the trajectories $(\mathbf{r}_i, t)$, then they also have to allow them to move on the trajectories expanded in time and space by a~constant factor, $\lambda(\vec{r}_i, t)$ where $\lambda$ is a~real number greater than 1. Consequently, the accelerations of bodies are enhanced compared to Newtonian dynamics in the low-acceleration, called the deep-MOND, regime.

All MOND theories are non-linear \citep{milgmondlaws}. This has the external field effect (EFE) as a~consequence: The internal dynamics of a~system is affected if the center of mass of the system is accelerating. In a~wide class of modified-gravity type MOND theories \citep{milgmondlaws}, the gravitational force of a~system in the deep-MOND gets weaker when the system is exposed to an external gravitational field. It differs from the tidal forces because the EFE occurs even for a~homogeneous external field or for arbitrarily small systems. In the case of the LG, the gravitational attraction between the MW and M\,31 comes out weaker if we take into account the external field caused by the nearby cosmic structures such as nearby galaxy clusters. With a~zero external field, the model by Z13 gives the MW-M\,31 pericenter around 7\,Gyr ago, and around 10\,Gyr ago with a~more realistic external-field strength of $0.03\,a_0$.

So far, two fully-fledged non-relativistic MOND theories were published both of which are modified gravity theories (AQUAL, \citealp{bm84}, and QUMOND, \citealp{qumond}). Their differences are still not explored well \citep{zhao10b}. For spherically symmetric matter density distributions, the theories give equal gravitational accelerations, but in other configurations, the accelerations can differ by tens of percent, see an example in \citet{banik15}. As \citet{candlish15} warns, the small difference in gravitational acceleration can accumulate over time and impact the evolution of the investigated system. Nevertheless, the existing comparative simulations seem to be little dependent of the particular MOND theory as far as we can judge from the figures in \citet{banik15, candlish15} or \citet{candlish16}.

Our simulations employed the QUMOND theory \citep{qumond}. For a~matter density distribution $\rho$, the QUMOND gravitational potential $\phi$ is given by the generalized Poisson equation
\begin{equation}
\Delta\phi= \vec{\nabla}\cdot\left[\nu\left(\left|\vec{\nabla}\phi_N\right|/a_0\right)\vec{\nabla}\phi_N\right],
\end{equation}
where $\phi_N$, determined by
\begin{equation}
\Delta\phi_N= 4\pi G \rho,
\end{equation}
is the Newtonian gravitational potential and $\nu$ is the interpolation function, in our simulations chosen as
\begin{equation}
\nu\left(y\right) = \left(1+\sqrt{1+4/y}\right)/2,
\end{equation}
which is known to produce a~good fit to galaxy rotation curves \citep{famaey05, gentile11} and for strong gravitational lenses \citep{sanders08}\footnote{However, see also \citet{hees16} for constraints in the Solar System, showing that this function, as well as many others, does not approach the Newtonian regime quickly enough in strong gravitational fields.}.
Point masses move in this gravitational field according to the usual equation of motion $\vec{\ddot r} = -\nabla \phi$.

We solved these equations with the publicly available Phantom of RAMSES adaptive-mesh-refinement code (PoR, \citealp{por}). The simulations contained only point masses in an invariable number. Cosmic expansion and an external field were not implemented. {The computational parameters of this simulation are listed in \tab{sim}. They lead to spatial resolution of 120\,pc.} {The PoR/RAMSES parameters not listed here were left default.}

\begin{table}
\caption{Simulation setup.}         
\centering                         
\begin{tabular}{ ll }\hline\hline
	Parameter & value\\ \hline
	$N(\,\mathrm{MW} \,)$ & $6\times 10^{4}$\\
	$N(\,\mathrm{M\,31} \,)$ & $16\times 10^{4}$\\
	\textsc{boxlen} & 4\,Mpc\\
	\textsc{levelmin} & 7\\
	\textsc{levelmax} & 15\\
 $\vec{r}_\mathrm{ini}(\, \mathrm{MW} \,)$ & $\left(\, -187.45,\, 41.875,\, -103.44 \,\right) $ kpc \\
	$\vec{v}_\mathrm{ini}(\, \mathrm{MW} \,)$ & $\left(\, 236.13,\, -13.745,\, 124.83 \,\right) $ km\,s$^{-1}$ \\
	$\vec{r}_\mathrm{ini}(\, \mathrm{M\,31} \,)$ & $\left(\, 70.351,\, -15.716,\, 38.820 \,\right) $ kpc \\
	$\vec{v}_\mathrm{ini}(\, \mathrm{M\,31} \,)$ & $\left(\, -88.406,\, 5.1461,\, -46.736 \,\right) $ km\,s$^{-1}$ \\
	\hline
\end{tabular}\label{tab:sim}
\tablefoot{$N(x)$ -- Initial number of particles in the galaxy $x$. \textsc{boxlen} -- Size of the computational cube. \textsc{levelmin} -- Minimum refinement level. \textsc{levelmax} -- Maximum refinement level. $\vec{r}_\mathrm{ini}(x)$ -- Initial position of the galaxy $x$. $\vec{v}_\mathrm{ini}(x)$ -- Initial velocity of the galaxy $x$. The PoR/RAMSES parameters not listed here were left default. }
\end{table}

\subsection{Choice of free physical parameters and coordinate system}\label{sec:param}
When choosing the physical parameters of our simulation, we were motivated by observations and, where applicable, we roughly followed the fiducial model by Z13. We used $a_0 = 1.2\times 10^{-10}$\,m\,s$^{-2}$. 

\begin{figure}[t!]
 \resizebox{\hsize}{!}{\includegraphics{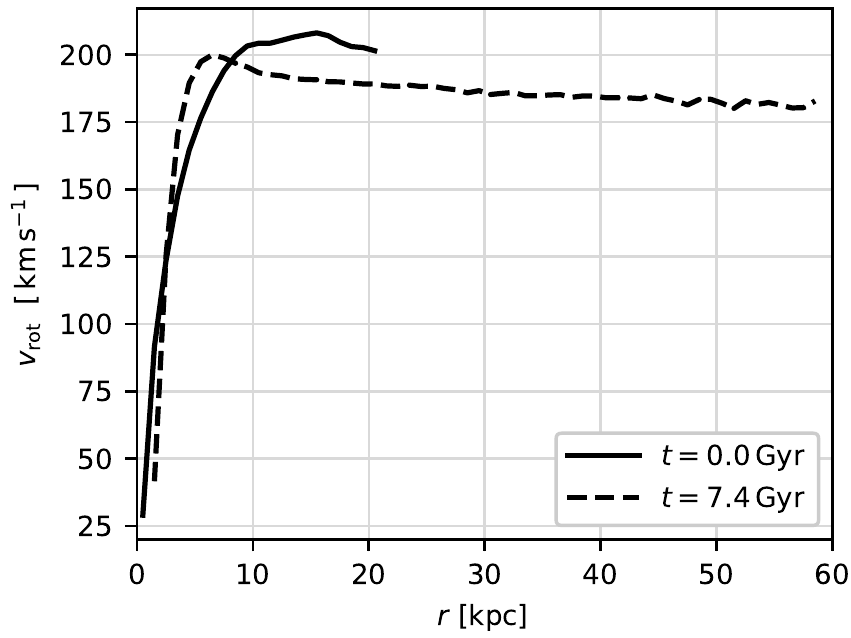}}
 \caption{{Rotation curve of the simulated MW at the simulation start (0.0\,Gyr) and at the current time (7.4\,Gyr). The model was initially truncated at 20\,kpc but the particles spread outwards as the galaxies developed bars and interacted such that the rotation could be traced to larger distances at the later times.}} 
 \label{fig:RCMW}
\end{figure}

\begin{figure}[h]
 \resizebox{\hsize}{!}{\includegraphics{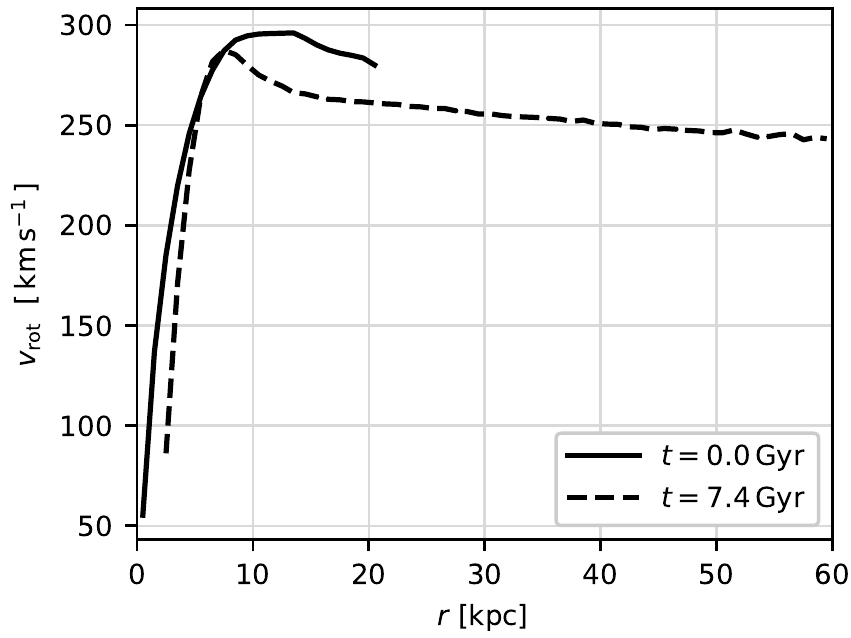}}
 \caption{{Rotation curve of the simulated M\,31 at the simulation start (0.0\,Gyr) and at the current time (7.4\,Gyr). The model was initially truncated at 20\,kpc but the particles spread outwards as the galaxies developed bars and interacted such that the rotation could be traced to larger distances at the later times.}} 
 \label{fig:RCM31}
\end{figure}

The galaxy masses were motivated by the baryonic Tully-Fisher relation, $V_\mathrm{f}^4 = GMa_0$, which is precise in all MOND theories. Here $V_\mathrm{f}$ is the rotational velocity at the asymptotically flat part of the rotational curve. This formula gives $M=6.6\times 10^{10}$\,M$_\sun$ for $V_\mathrm{f} = 180$\,km\,s$^{-1}$ of the MW (Fig.~4 of \citealp{wu08}, following Z13) and $M=1.6\times 10^{11}$\,M$_\sun$ for $V_\mathrm{f} = 225$\,km\,s$^{-1}$ of M\,31 \citep{carignan06}. In the simulation, we used 
\begin{equation}
M_\mathrm{MW} = 6.14\times 10^{10}\,\mathrm{M}_\sun
\label{eq:mmw}
\end{equation}
and
\begin{equation}
M_\mathrm{M\,31} = 16.36\times 10^{10}\,\mathrm{M}_\sun.
\label{eq:mm31}
\end{equation}
The galaxies were modeled as truncated exponential disks with a~density distribution of the form 
\begin{equation}
\rho(r, z) = \rho_0\,\exp\left(-r/r_\mathrm{d}\right)\mathrm{sech}^2\left(z/z_0\right),
\label{eq:disk}
\end{equation}
for $r\leq r_\mathrm{t}$ and $\rho(r, z) = 0$ for $r>r_\mathrm{t}$ with the truncation radius $r_\mathrm{t}$ of 20\,kpc. The central density $\rho_0$ was scaled to obtain the required disk masses. The real scale length of the MW is around 2.1\,kpc \citep{bovy13} and that of M\,31 is around 5.3\,kpc \citep{courteau11}. In the simulation, we implemented the scale length $r_\mathrm{d}$ of 3.5\,kpc and the scale height $z_0$ of 0.3\,kpc for both galaxies (i.e. the half-mass radius, $r_\mathrm{h} = 1.67\,r_\mathrm{d}$, of 5.8\,kpc). The velocity dispersion was set to get the Toomre Q parameter of 1 at all radii according to the Eqs.~20-22 of \citet{por}. The initial models for both galaxies were set up by the method and code described in Sect.~5.1 of \citet{por}. We show how our galaxy models evolve in isolation in Appendix~\ref{app:app}. In short, they quickly develop bars and increase the effective radii by at most 25\% between the simulation start and the time when the observed separation and relative velocity were reproduced. {Figures~\ref{fig:RCMW} and~\ref{fig:RCM31} show the rotation curves of the simulated MW and M\,31. }

We used the same axes directions as \citet{vandermarel12} who measured the proper motion of M\,31 using the Hubble Space Telescope: The $Z$ axis pointed from the observed MW center to its northern pole; the $X$ axis pointed from the current observed position of Sun to the MW center; and $Y$ axis pointed in the direction of the Sun motion around the MW center. We adopted the MW-M\,31 barycenter as the origin of our coordinate system. Then, following \citet{vandermarel12}, the vector from the MW center to the M\,31 center is (their Eq.~2)
\begin{equation}
\vec{r}_{\mathrm{MW}-\mathrm{M\,}31} = \left(\,-378.9,\, 612.7,\,-283.1\,\right) \mathrm{\,kpc},
\label{eq:r}
\end{equation}
and the most probable relative velocity of M\,31 with respect to MW measured by \citet{vandermarel12} is (their Eq.~3)
\begin{equation}
\vec{v}_{\mathrm{MW}-\mathrm{M\,}31} = \left(\,66.1,\,-76.3,\, 45.1\,\right) \mathrm{\,km\,s}^{-1}.
\label{eq:v}
\end{equation}
We adopted the current spin directions of the disks from \citet{pawlowski13} who used the same axes directions
\begin{equation}
\vec{s}_{\mathrm{MW}} = \left(\,0,\, 0,\, -1\,\right) \mathrm{\,kpc},
\label{eq:smw}
\end{equation}
\begin{equation}
\vec{s}_{\mathrm{M\,31}} = \left(\,-0.420,\, -0.757,\, -0.500\,\right) \mathrm{\,kpc}.
\label{eq:sm31}
\end{equation}
The position of the Sun in our coordinate system was 
\begin{equation}
\vec{r}_\odot = \vec{r}_\mathrm{MW} - \left(\,8.5,\, 0,\, 0 \,\right) \mathrm{\,kpc},
\end{equation}
 where we assumed the same distance of the Sun from the MW center as \citet{vandermarel12}. 

We assumed a~zero external field, i.e. the LG in our simulation is not subject to an EFE from neighboring objects.

None of these parameters was tuned to achieve the results described in \sect{res}.

\begin{figure}
 \resizebox{\hsize}{!}{\includegraphics{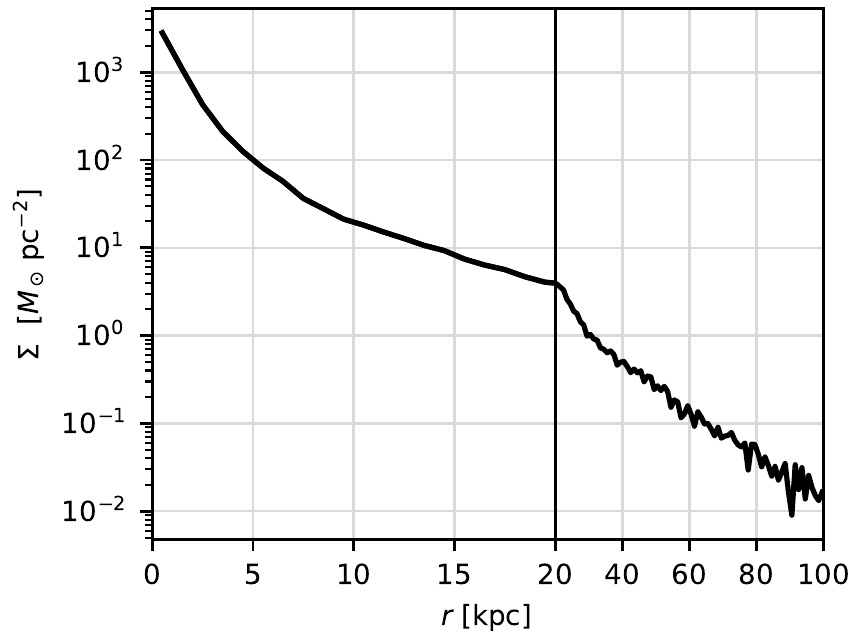}}
 \caption{{Surface density profile of the simulated MW at the current time.}} 
 \label{fig:SDMW}
\end{figure}

\begin{figure}
 \resizebox{\hsize}{!}{\includegraphics{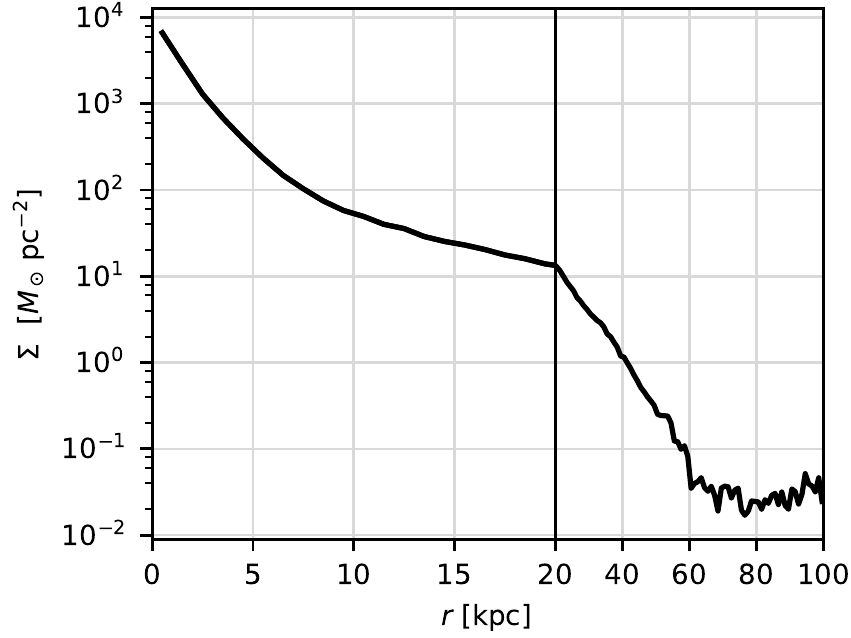}}
 \caption{{Surface density profile of the simulated M\,31 at the current time.}} 
 \label{fig:SDM31}
\end{figure}

\subsection{Defining the basic galaxy properties in the simulation}
To obtain the position and velocity of a~galaxy in the simulation, we proceeded in the following way. Each particle had an identifier assigned according to its parent galaxy. To define a~position of the galaxy $i$ in the later time steps, we applied 60 passes of the sigma-clipping algorithm to the particles that originally belonged to the galaxy $i$. This means: 1) calculate the center of mass of the particles in consideration, $\vec{x}$, and the root-mean-square of the particle distances from $\vec{x}$, $\sigma$; 2) for the next iteration, consider the particles whose distance from $\vec{x}$ is lower than $3\sigma$. These steps are repeated iteratively. The velocity of a~galaxy is calculated as the average velocity of the particles considered in the last iteration of the sigma-clipping algorithm.

We defined the galaxy spins in the simulation using the principal component method applied to the positions of the particles closer than 15\,kpc from the respective galaxy center and chose the orientation according to the galaxy rotation sense. The disk midplanes of the galaxies were considered being perpendicular to the spins and {going through} the galaxy positions.

{With} particles belonging to a~galaxy, we mean the particles that were closer to it than to the other galaxy at the given simulation time.


\subsection{Finding the orbital parameters}
\label{sec:orbit}
We aimed to find initial conditions so that the MW and M\,31 in the simulation reproduce the observed MW-M\,31 distance, relative velocity and disk inclinations at some simulation time.

To get the initial positions, we integrated the motion of the MW and M\,31 analytically backwards starting from the observed state (\equ{r} and \equ{v}) in the coordinate system defined in \sect{param}. We modified the prescription for the two-body force in MOND \citep{milg94c, zhao10}
\begin{equation}
F=\frac{Gm_1m_2}{r^2}+\frac{\Xi\sqrt{G\left(m_1+m_2\right)^3a_0}}{r},
\label{eq:tbf}
\end{equation}
\begin{equation}
\Xi\equiv \frac{2}{3}\left[1-\sum_{i = 1}^2\left(\frac{m_i}{m_1+m_2}\right)^{3/2}\right],
\end{equation}
 to account for the internal sizes of the galaxies by replacing the distance between the point masses $r$ in \equ{tbf} by $\sqrt{r^2+b_1^2+b_2^2}$. Here $b_i = 1.28r_{\mathrm{d},i}$ is the Plummer radius of a~Plummer sphere having the same half-mass radius as an exponential disk with the scale length $r_{\mathrm{d},i}$ (in our case, $r_{\mathrm{d},1}=r_{\mathrm{d},2} = 3.5$\,kpc, see \sect{param}). We integrated the motion of the galaxies backwards until they reached the pericenter and then receded to 300\,kpc from each other.

\begin{table}
\caption{Comparison of the real and simulated LG properties.}     
\centering 
\begin{tabular}{ lrr }\hline\hline
 Quantity & Real  & Simulated \\\hline
$\angle(\, \vec{s}_\mathrm{MW}, \vec{r}_{12} \,) ~~ [\, \degr \,] $ & 69 & 79 \\
$\angle(\, \vec{s}_\mathrm{MW}, \vec{v}_{12}\,) ~~ [\, \degr \,]$ & 114 & 104 \\
$\angle(\, \vec{r}_{12}, \vec{v}_{12}\,) ~~ [\, \degr \,]$ & 171 & 171 \\
$\angle(\, \vec{s}_\mathrm{MW}, \vec{s}_\mathrm{M\,31} \,) ~~ [\, \degr \,] $ & 60 & 76 \\
$\angle(\, \vec{r}_{12}, \vec{s}_\mathrm{M\,31} \,) ~~ [\, \degr \,]$ & 102 & 98 \\
$v_\mathrm{r} ~~ [\, \mathrm{km}\,\mathrm{s}^{-1}\, ]$ & -109.3 & -109.9 \\
$v_\mathrm{t} ~~ [\, \mathrm{km}\,\mathrm{s}^{-1}\, ]$ & 17.0 & 18.3 \\
$r_{12} ~~ [\,\mathrm{kpc}\,]$ & 774.0 & 774.3 \\
$r_{\mathrm{h},\mathrm{MW}} ~~ [\, \mathrm{kpc} \,]$ & 3.5 & 4.0 \\
$r_{\mathrm{h},\mathrm{M\,31}} ~~ [\, \mathrm{kpc} \,]$ & 8.9 & 4.2 \\\hline
\end{tabular}\label{tab:comp}
\tablefoot{ {The real values mean the observational values adopted in this paper}. The simulated values refer to the current time. $\angle(\vec{x}, \vec{y})$ -- Angle between the vectors $\vec{x}$ and $\vec{y}$.  $\vec{s}_x$ -- Spin vector of the galaxy $x$. $\vec{v}_{12}$ -- MW and M\,31 relative velocity vector. $v_\mathrm{r}$ -- Relative radial velocity. $v_\mathrm{t}$ -- Relative tangential velocity. $r_{12}$ -- Relative distance. $r_{\mathrm{h},x}$ -- Half-mass radius of the galaxy $x$.}
\end{table}

\begin{table}
\caption{Important orbital events in the simulated LG.}     
\centering
\begin{tabular}{ lrcc }\hline\hline
	Event & $t$ [Myr] & $r_{12}$ [kpc] & $v_{12}$ [km\,s$^{-1}$] \\\hline
 Simulation start & 0 & 300 & 368\\
	First pericenter & 653 & 24.2 & 637\\
	First apocenter & 5491 & 880 & 17.0 \\
	Current time & 7411 & 774 & 111\\
	Second pericenter & 10478 & 21.5 & 608\\
	Second apocenter & 13953 & 627 & 18.0\\
	Simulation end & 14017 & 626 & 17.4\\	\hline
\end{tabular}\label{tab:events}
\tablefoot{$t$ -- Simulation time. $r_{12}$ -- Galaxy separation. $v_{12}$ -- Galaxy relative velocity magnitude.}
\end{table}

The initial spin vectors of the simulated galaxies were chosen as the observed spin vectors (\equ{smw} and \equ{sm31}): We assumed that the spins would not change much by the encounter and that the MW-M\,31 orbit in the self-consistent simulation would not change substantially by dynamical friction compared to the analytic orbit. This is assessed in \tab{comp}.

Then we{ searched} for the initial velocities required to reproduce the observed relative MW-M\,31 velocity and separation (but not necessarily the observed positions given by \equ{r}). We proceeded iteratively. In the first iteration, the galaxies had their initial velocities taken from the analytical orbit. {We then} ran a~self-consistent simulation while checking its output every 20\,Myr. When the galaxies reached the apocenter and {when} their separation dropped under the observed value, the simulation was stopped and the relative galaxy radial and tangential velocities were compared to their observed values. Then we adjusted the initial velocities for the next iteration by changing the initial relative radial and tangential velocity magnitudes (i.e. the plane of the encounter stayed fixed). This was repeated until both the final relative radial and tangential velocities differed by less than 3\,km\,s$^{-1}$ from the observed values. {The final simulation was continued to cover {a~time} of 14\,Gyr.}

The complete setup of the final simulation is summarized in \tab{sim}.

\begin{figure}[t!]
 \resizebox{\hsize}{!}{\includegraphics{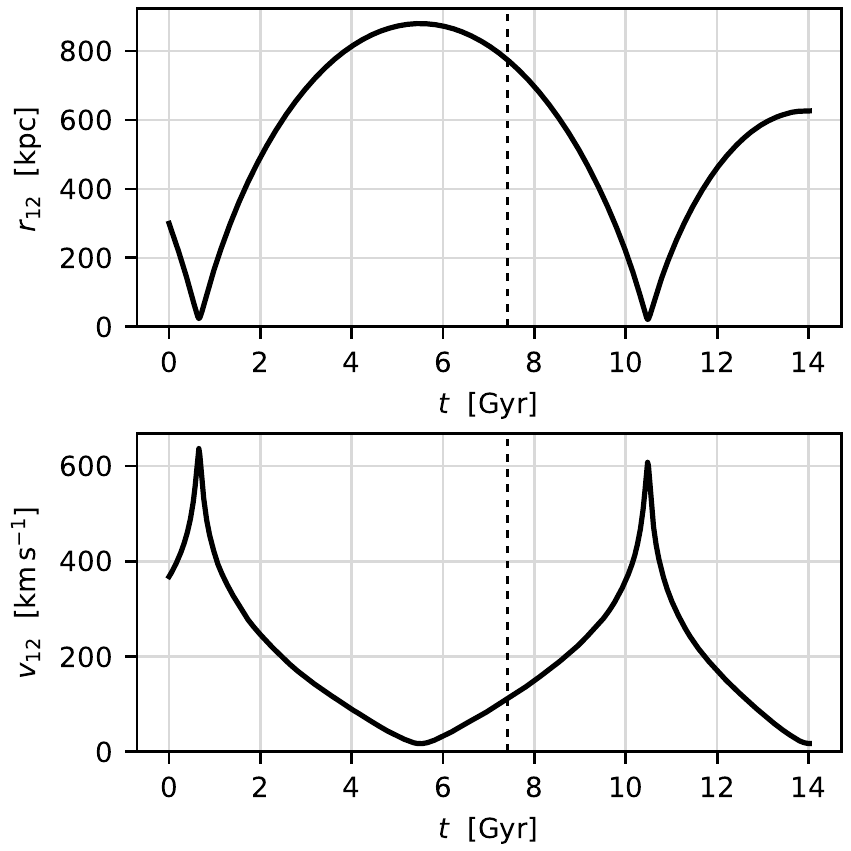}}
 \caption{Top -- Evolution of galaxy separation, $r_{12}$, with the simulation time, $t$. Bottom -- Evolution of galaxy relative velocity magnitude, $v_{12}$, with the simulation time. The vertical dashed line indicates the current time (7411\,Myr).}
 \label{fig:trv}
\end{figure}

\begin{figure}[h!]
 \resizebox{\hsize}{!}{\includegraphics{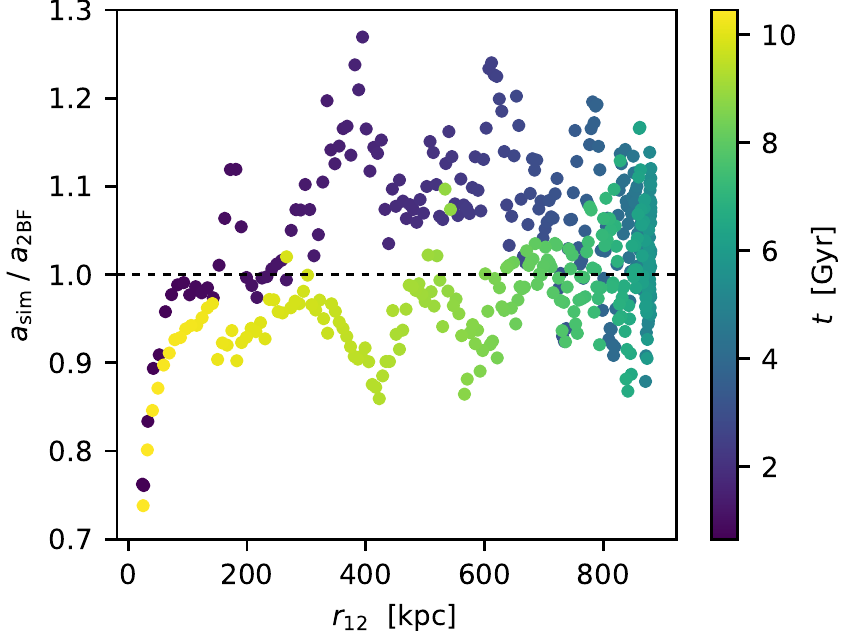}}
 \caption{Ratio of the relative accelerations that the galaxies have in the simulation, $a_\mathrm{sim}$, to the acceleration calculated using the two-body-force formula (\equ{tbf}), $a_\mathrm{2BF}$, as a~function of the galaxy separation, $r_{12}$. If the accelerations $a_\mathrm{sim}$ and $a_\mathrm{2BF}$ were equal, the points would lie on the horizontal dashed line. The points are colored with respect to simulation time, $t$. Only the period between the first and second pericenter is displayed.} 
 \label{fig:acccomp}
\end{figure}


\begin{figure*}
\centering
  \includegraphics[scale = 1]{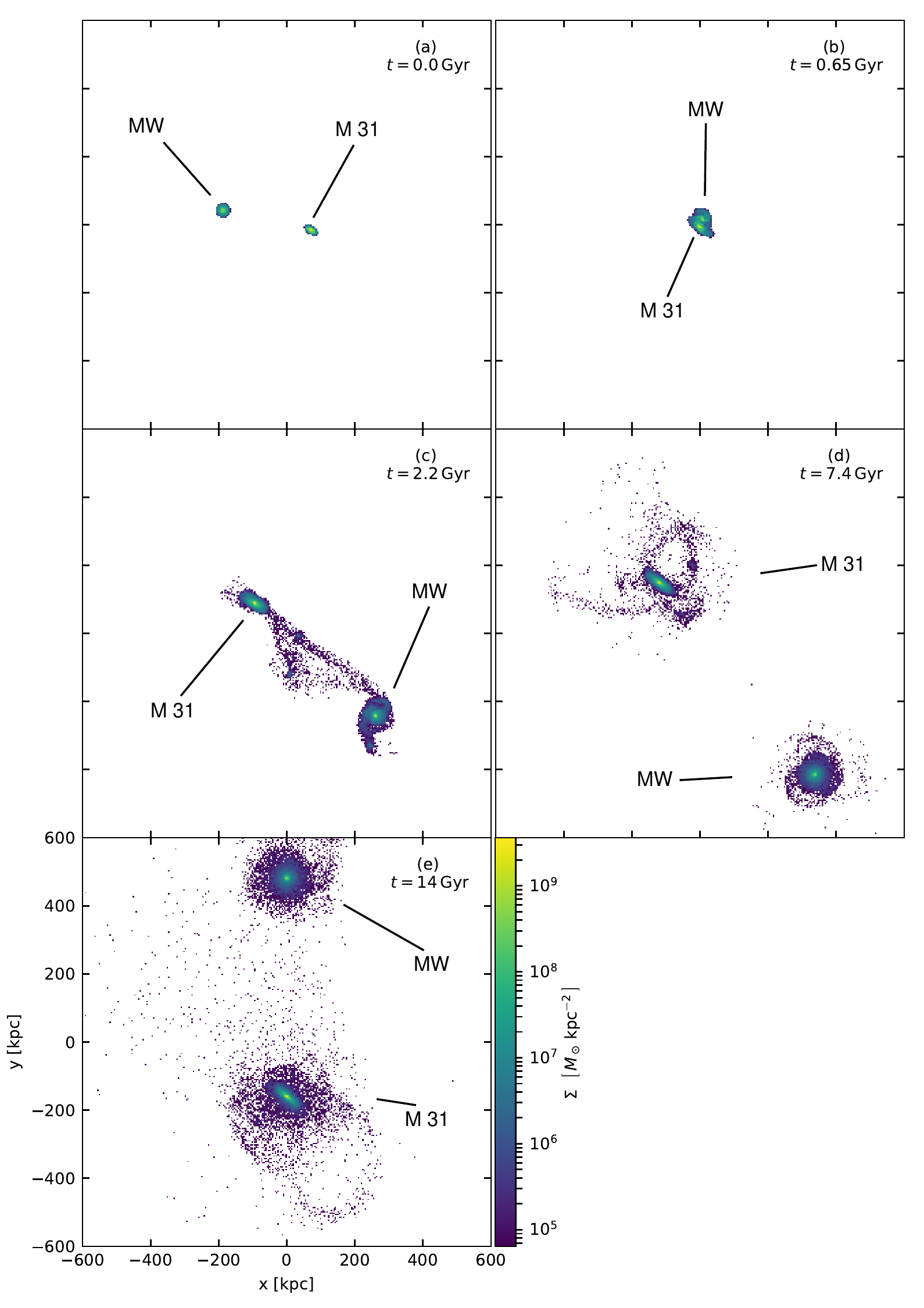}
   \caption{{Snapshots of the simulation. Projection along the $Z$-axis. The time since the beginning of the simulation is marked. \textit{(a)} Simulation starts. \textit{(b)} Galaxies are in the relative pericenter. \textit{(c)} Matter is being transferred from the MW to M\,31. \textit{(d)} The current time (the observed state is reproduced). \textit{(e)} Simulation ends.}}
   \label{fig:simall}
\end{figure*}

\section{Results}\label{sec:res}

As {described} above, our simulation was set so that it approximately reproduces the observed MW-M\,31 separation, radial and tangential velocity, disk spin direction, masses and scale lengths. {What follows in this section is the consequence of this setup.}

We denote by the term ``the current time'' the moment when the simulated MW and M\,31 reached the observed separation for the first time after coming through their first relative apocenter (and also the observed radial and tangential velocities are reproduced by design, see \sect{orbit}). The current time occurred 7411\,Myr after the simulation start. In \tab{comp}, we compare the above mentioned quantities at the current time to the adopted real values.

\begin{figure}
 \resizebox{\hsize}{!}{\includegraphics{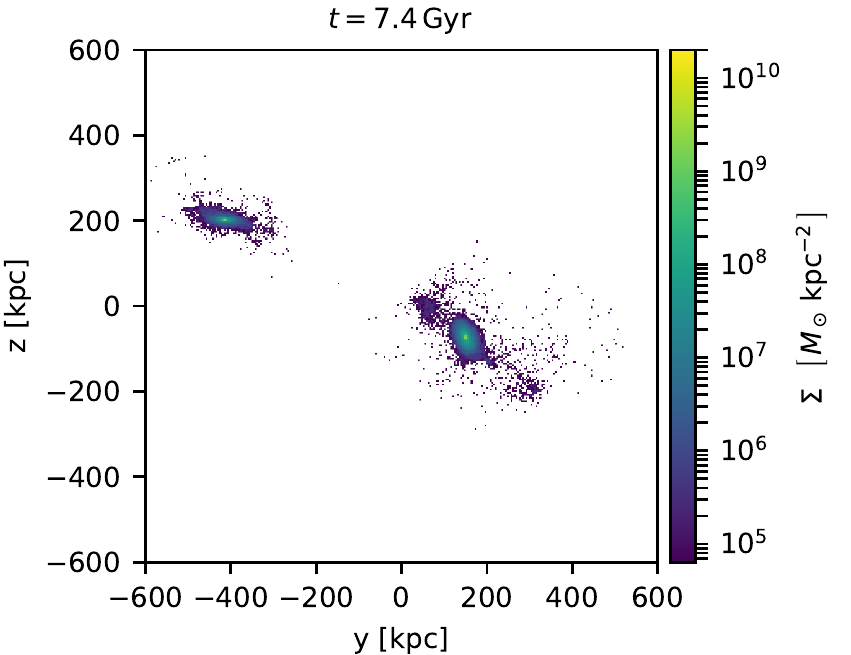}}
 \caption{Current time. Projection along the $X$-axis. Simulation time of 7.4\,Gyr. The MW is on the left, M\,31 on the right.} 
 \label{fig:yzproj}
\end{figure}

\begin{figure}
 \resizebox{\hsize}{!}{\includegraphics{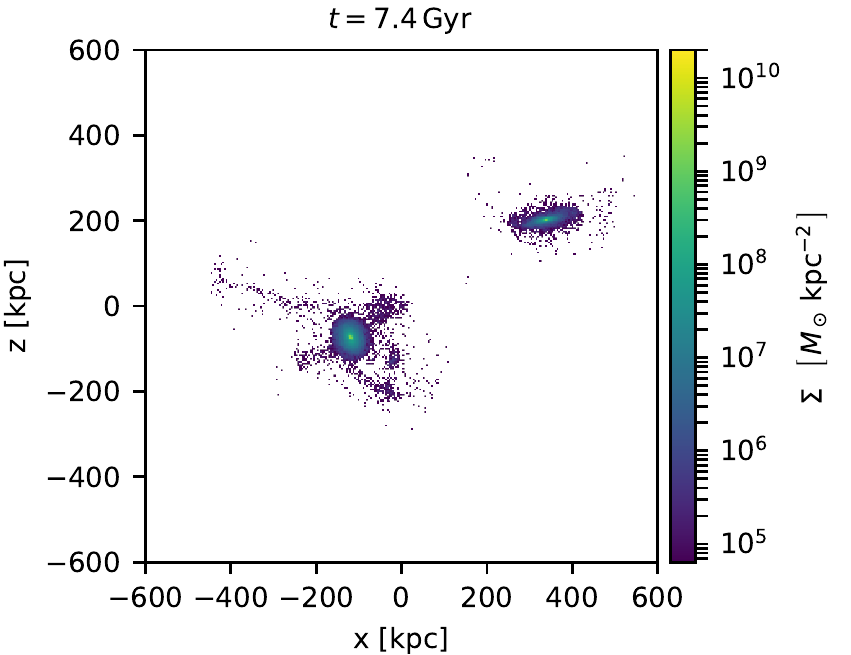}}
 \caption{Current time. Projection along the $Y$-axis. Simulation time of 7.4\,Gyr. The MW is on the right, M\,31 on the left.} 
 \label{fig:xzproj}
\end{figure}

\begin{figure}
 \resizebox{\hsize}{!}{\includegraphics{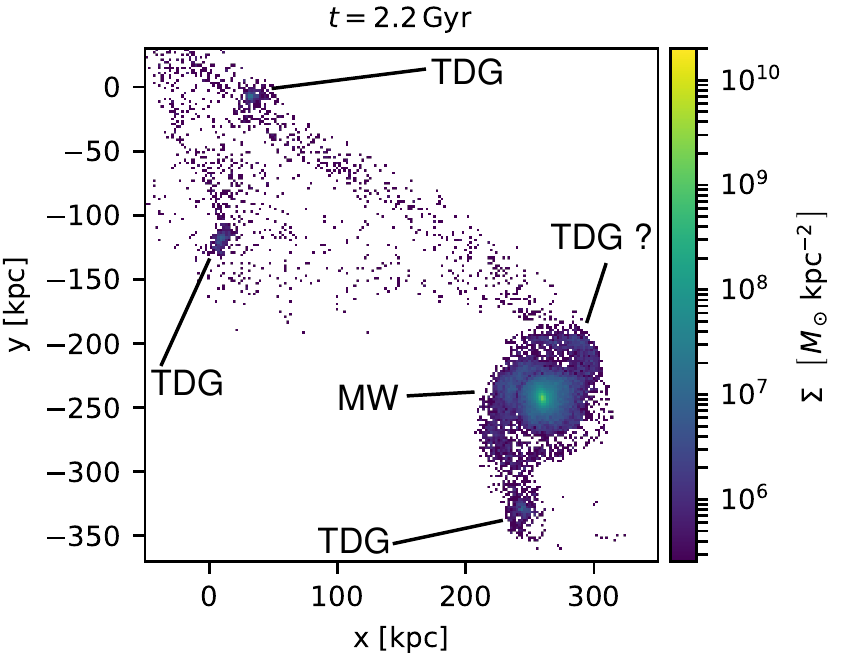}}
 \caption{{Zoom at the tidal dwarf galaxies (TDGs) in the simulation at the simulation time of 2.2\,Gyr. }} 
 \label{fig:tdgzoom}
\end{figure}

\begin{figure}
 \resizebox{\hsize}{!}{\includegraphics{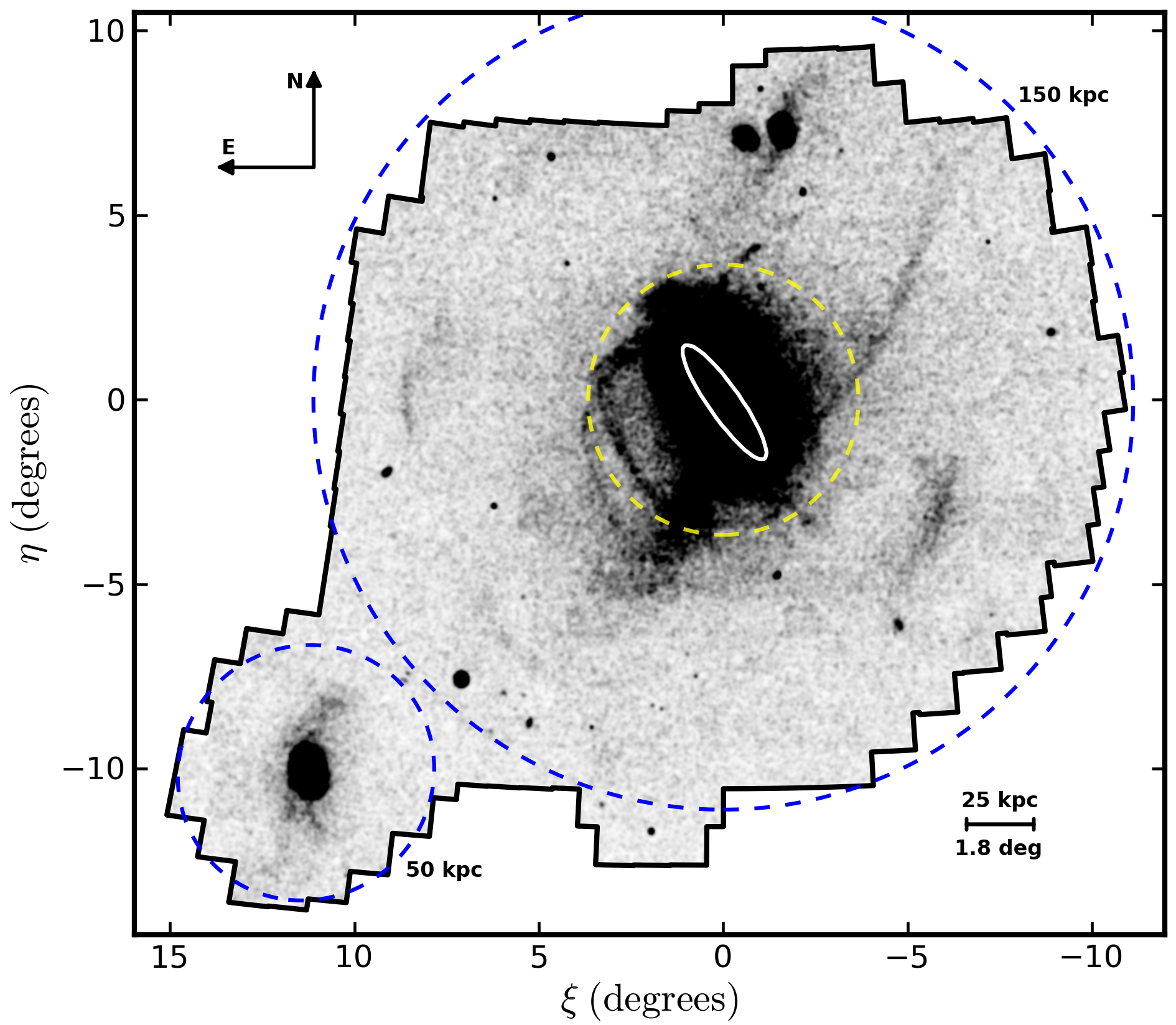}}
 \caption{Stellar streams around M\,31 as observed by the PAndAS survey. Note the similarity to the tidal features around the simulated M\,31 in Figs.~\ref{fig:simall} (d) and~\ref{fig:xzproj}.  Image courtesy Dougal Mackey (adapted Fig.~3 from \citealp{ferguson16}).}
 \label{fig:streams}
\end{figure}

The time evolution of the galaxy separation and relative velocity magnitude are drawn in \fig{trv}. The vertical dashed line indicates the current time. The first pericentric passage in our simulation occurred 6.8\,Gyr before the current time when the galaxy separation was 24\,kpc and the relative velocity reached 637\,km\,s$^{-1}$. To make a~comparison to the method used by Z13, we integrated analytically the orbit backward using the two-body-force formula \equ{tbf} and the galaxy masses as in our simulation. This resulted in the pericentric passage closer to today surprisingly only by 0.06\,Gyr compared to the self-consistent simulation. The next pericentric passage occurs in the self-consistent simulation 3.1\,Gyr after the current time. For comparison, the \lcdm simulation by \citet{vandermarel12future} gives the first future pericentric approach of the MW and M\,31 in around 3.9\,Gyr and the final merger in 5.9\,Gyr from now. The galaxies were still far from merging at that time in our MOND simulation{ because of the reduced dynamical friction because the particle dark matter halos do not occur}. The times of the important orbital events in the simulation are listed in \tab{events} along with the respective galaxy separations and relative velocity magnitudes. We remind that these times would change significantly if the EFE and the cosmic expansion were taken into account (see Z13).

\begin{figure}
 \resizebox{\hsize}{!}{\includegraphics{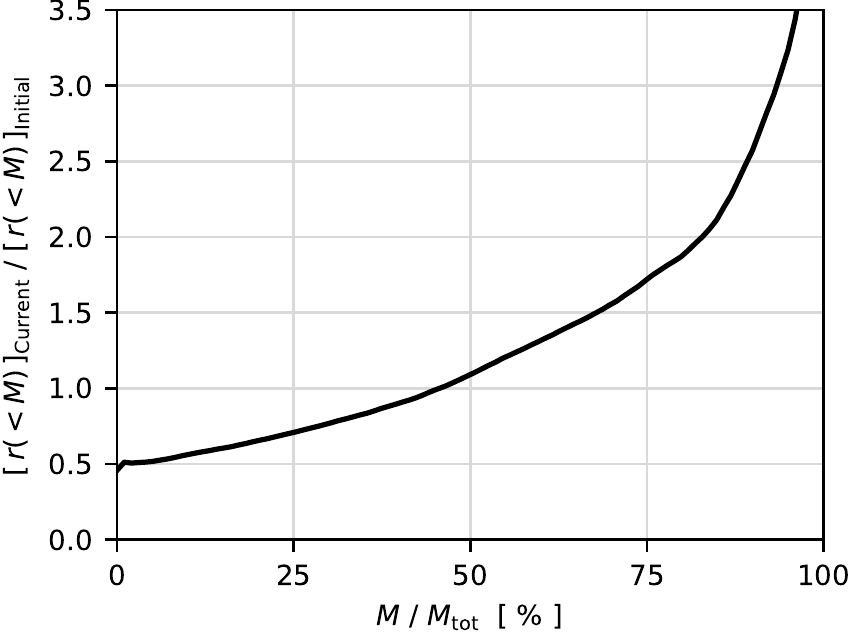}}
 \caption{Ratio of the radius enclosing the mass on the horizontal axis at the current time to the radius enclosing the same mass at the simulation start for the MW.} 
 \label{fig:mwlag}
\end{figure}

\begin{figure}
 \resizebox{\hsize}{!}{\includegraphics{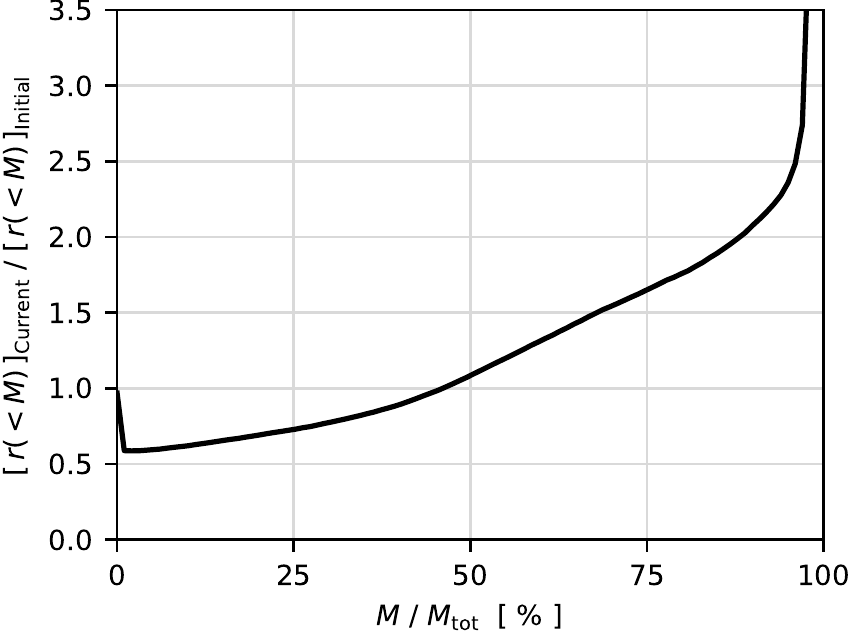}}
 \caption{Ratio of the radius enclosing the mass on the horizontal axis at the current time to the radius enclosing the same mass at the simulation start for the M\,31.} 
 \label{fig:m31lag}
\end{figure}

We plotted in \fig{acccomp} the ratio of the relative acceleration magnitude {of the galaxies measured from their} velocity change in the simulation to the acceleration calculated using the two-body-force formula \equ{tbf} as a~function of the galaxy separation color-coded according to the simulation time. Only the period between the first and second pericentric passage was plotted. The initial galaxy masses were assumed for the analytic calculation. {Here we noted several unexpected facts: 1) the acceleration ratio was different for the receding (blue) and approaching (green and yellow) part of the orbit, 2) the major wiggles up in the receding part were followed by similar wiggles down in the approaching part when the galaxies had the same separation, and 3) the acceleration ratio evolved in time even in the apocenter when the separation was changing only little. We left the explanation of these effects for future investigations.}

Figure~\ref{fig:simall} shows a~few snapshots from our simulation at different simulation times. The videos showing the simulation projected along the $X, Y$ and $Z$ axes are available online\footnote{The videos from our simulation are available at this url: \url{http://galaxy.asu.cas.cz/~bilek/LGindex.html}.}. The projections of the simulation along the $X, Y$ and $Z$ axes at the current time show Figs.~\ref{fig:yzproj}, \ref{fig:xzproj} and \ref{fig:simall} (d), respectively.  It might appear from these plots that the simulated galaxies are too big at the current time compared to the real MW and M\,31, but this is an effect of the used logarithmic color scale. The surface density profiles of the simulated galaxies at the current time are shown in Figs.~\ref{fig:SDMW} and~\ref{fig:SDM31}. A~comparison with the second row of Fig.~5 by \citet{yin09} shows that the surface density profiles are correct within an order of magnitude. {Whether the real MW disk has a~break beyond around 12\,kpc is a~matter of an ongoing debate \citep{bland16, carraro15}.}

We can see that the encounter induced formation of tidal tails in the MW (see \citealp{bournaud10} for an explanation of the tidal tail formation mechanism). One of the tidal tails in our simulation was ejected from the MW towards M\,31 where its part was captured. The remaining part was captured by the MW. This formed tidal {structures} visible in the simulation at the current time around both galaxies. The bridge connecting the MW and M\,31 has been visible for around 4\,Gyr. The other tidal tail was captured by the MW completely. There were no particles receding from the LG barycenter {by} more than a~few hundreds of kiloparsecs (actually, there could be no escaping particles because the escape speed from isolated objects in MOND is infinite). Only minor hints of tidal tails were induced in M\,31. Defining the transferred matter as the particles that belonged to one galaxy at the beginning of the simulation and belonged to the other galaxy at the current time, 3.2\% of MW mass was transferred to M\,31 and no mass was transferred in the opposite direction.

{Three or four temporal TDGs could be visually detected to form in the tidal tails in the simulation (\fig{tdgzoom} is a~magnified version of \fig{simall} (c) providing a~detailed view of the TDGs). { Note that the number of TDGs formed in the simulation would most probably increase if gas and star formation were included since the gas cooling facilitates the formation of gravitationally bound objects. The two TDGs in the} tail pointing to M\,31 were captured to become satellites of M\,31. One or two TDGs formed in the tidal arm pointing initially away from M\,31 and stayed bound to the MW. This simulation therefore confirms the earlier finding that TDGs form easily during galaxy interactions in MOND. However, \citet{wetzstein07} found, in Newtonian gravity, that the formation of the
TDGs should be studied only in simulations containing gas (see \sect{disc} for more details). Nevertheless, MOND simulations with gas indeed show TDGs forming in tidal tails (see, e.g., Figs.~2 and~3 in \citealp{renaud16}), so that the MW and M\,31 could be enriched by new satellites in the way suggested by our simulation.  The relation of the satellites to the tidal streams here is partly opposite from the classical view: Here the streams existed first and collapsed temporarily into the TDGs before the TDGs were disrupted into streams again when the tidal forces and the EFE increased (the satellite disruption in MOND was detailed by \citealp{satelliteefe}).}

%
%
%

To assess how the galaxy radial profiles, including the half-mass radii, evolved from the initial state to the current time, we made Figs.~\ref{fig:mwlag} and \ref{fig:m31lag} for the MW and for M\,31, respectively. These plots show the ratio of the Lagrangian radius for the current time to the Lagrangian radius at the simulation start for every mass on the horizontal axis for the respective galaxy (a Lagrangian radius is the radius enclosing a~given mass). These plots imply for both galaxies that the half-mass radii almost did not change, while the galaxy centers shrunk and the outer parts expanded. This might be related to bulge formation.

{The rotation curves of the MW and M\,31 at the beginning of the simulation and at the current time are displayed in Figs.~\ref{fig:RCMW} and~\ref{fig:RCM31}. To obtain the rotation velocity $v$ at some galactocentric radius $r$, we chose all particles with a~galactocentric distance  between $r-0.5$\,kpc and $r+0.5$\,kpc and with a~distance from the galaxy midplane less than 0.5\,kpc. Then we calculated their average radial acceleration $a_\mathrm{rad}$ from their velocity change between the subsequent time  steps and calculated the rotational speed $v = \sqrt{a_\mathrm{rad}\, r}$. The rotation curves can be compared to observations in \citet{bland16} for the MW and in \citet{carignan06} for M\,31. The difference is typically 10-20\%.}

For what follows, we needed to define in the simulation at the current time the position and velocity of the Sun and analogues of the equatorial and the Galactic coordinate systems. The Sun lied at the point meeting the conditions: 1) It lied in the MW midplane, 2) Its distance from the MW center was 8.5\,kpc, 3) In the obvious analogue of the Galactic coordinate system in the simulation, M\,31 had the observed Galactic longitude. The MW north pole direction in the simulation was defined as the opposite vector to the MW spin, just as in the real MW. We defined the equatorial coordinate system connected with the Sun so that the MW center and its pole had the observed coordinates in it.  We assumed that the Sun's orbital velocity vector lied in the MW disk midplane, was perpendicular to the direction to the MW center and agreed with the net rotation sense of the galaxy.

\begin{figure}
 \resizebox{\hsize}{!}{\includegraphics{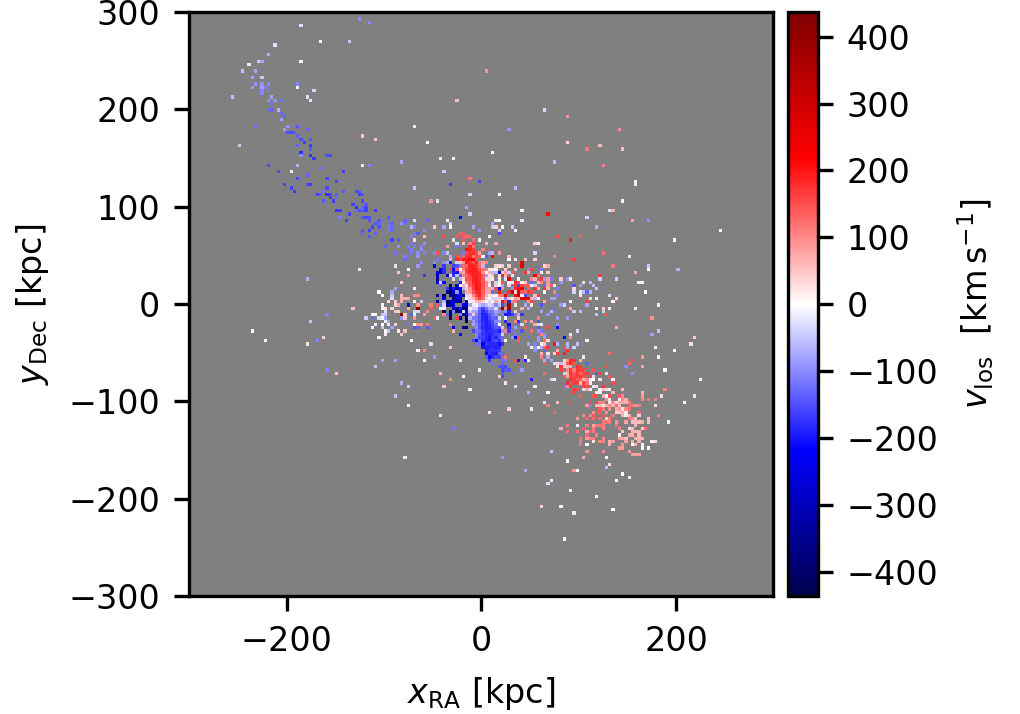}}
 \caption{View of M\,31 from the Sun in the simulation. The color codes the average line-of-sight velocity with the systemic velocity subtracted. Note the pronounced linear feature resembling the GPoA (\fig{m31sp}) and the dissolving satellite in the right part (better visible in \fig{xzproj}) similar to the dissolving satellite in \fig{streams}.} 
 \label{fig:m31sun}
\end{figure}

\begin{figure}
 \resizebox{\hsize}{!}{\includegraphics{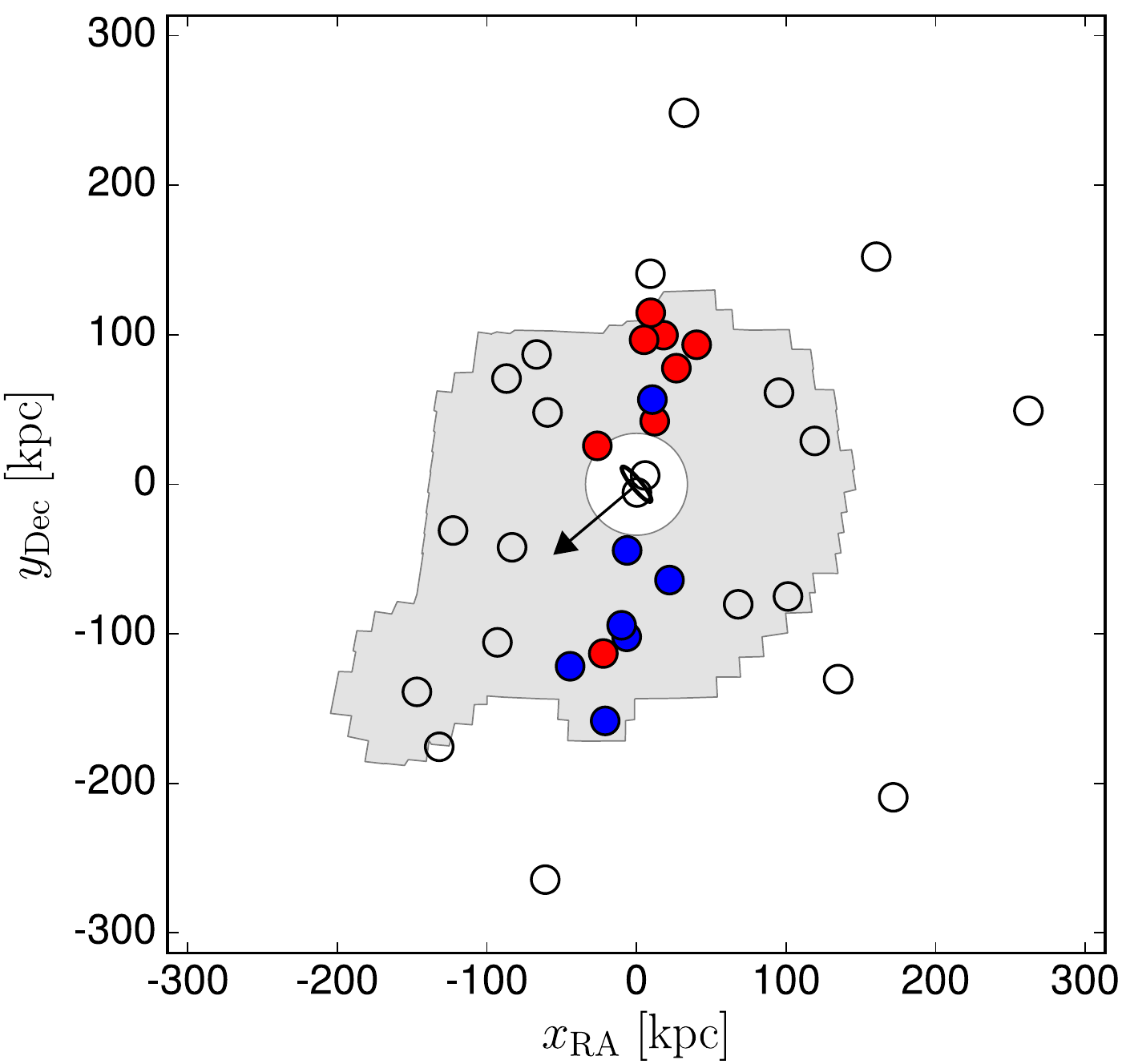}}
 \caption{The satellites of M\,31 (circles). The blue and red satellites belong to the GPoA. The satellites in blue are approaching toward the Earth from the coordinate system connected with M\,31, the satellites in red are receding. The arrow indicates the spin of the M\,31 galactic disk. Compare to the model shown in \fig{m31sun}. The area covered by the PAndAS survey is shown in gray to facilitate the comparison with the image of the galaxy \fig{streams}. Image courtesy Marcel Pawlowski (adapted Fig.~11 from \citealp{bullock17}).} 
 \label{fig:m31sp}
\end{figure}

\begin{figure}
 \resizebox{\hsize}{!}{\includegraphics{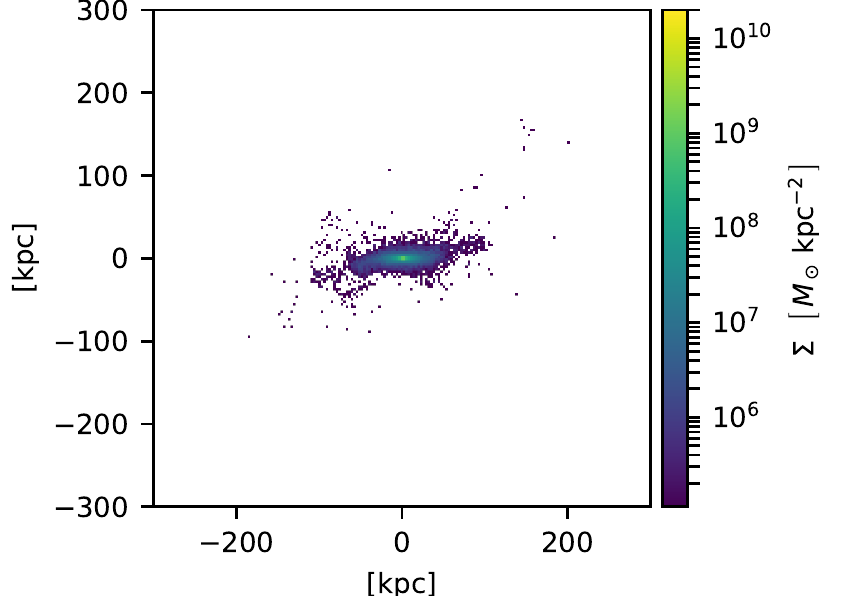}}
 \caption{Cloud of particles formed around the simulated MW. The SP candidate is seen edge-on here. {A~SP that would be perpendicular to the MW galactic disk, just as the VPOS }(see, e.g., Fig.~2 of \citealp{kroupacjp}), was not formed in this simulation, but the extent of the particle cloud here matches well the extent of the VPOS.} 
 \label{fig:mwdosedge}
\end{figure}

\begin{figure*}[t!]
\centering
  \includegraphics[width=17cm]{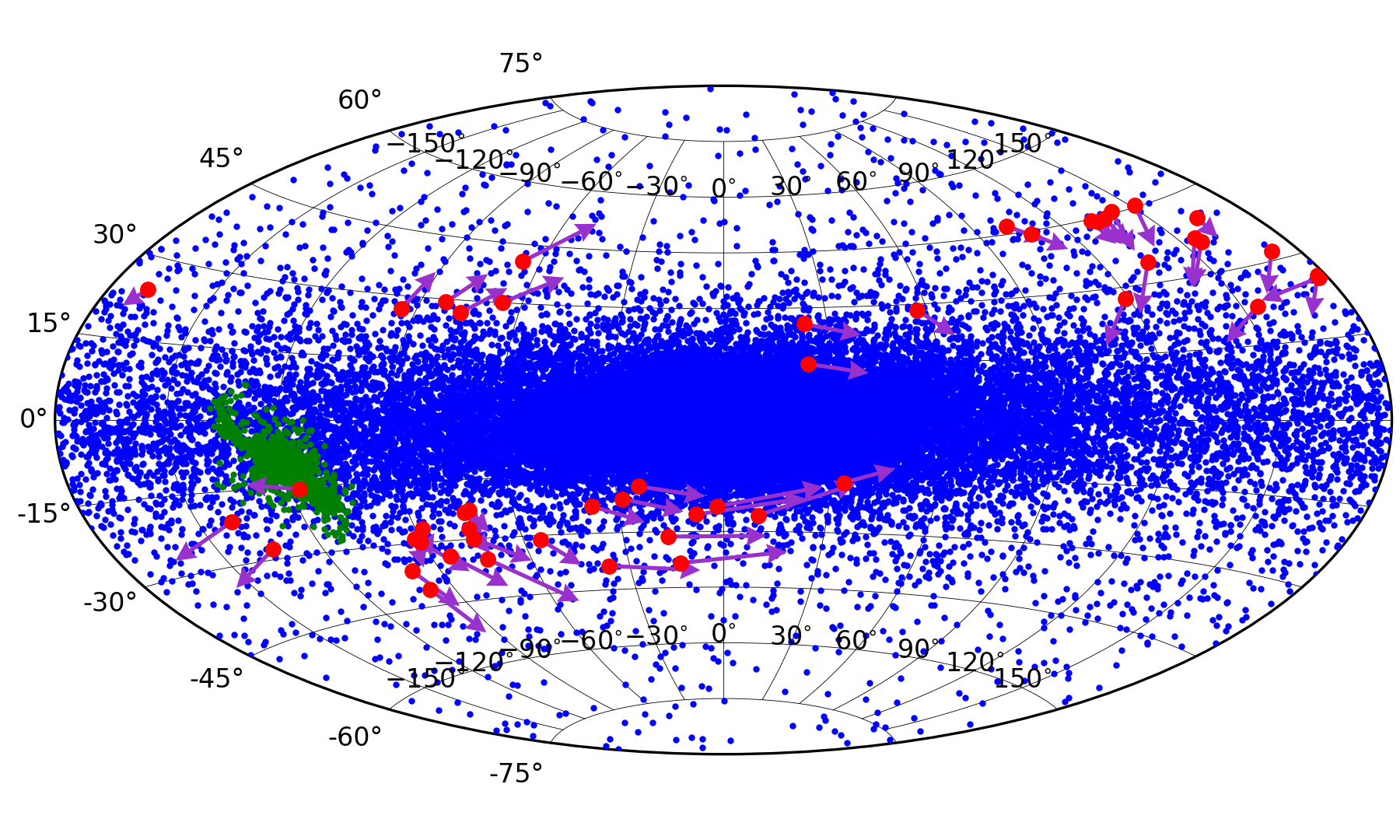}
   \caption{Aitoff projection of all particles in the simulation seen from the position of the Sun in the Galactic coordinate system. The M\,31 particles are green. The particles belonging to MW further than 50\,kpc from the MW disk plane are red. The other MW particles are blue. The arrows show the proper motions of the red particles as observed from the Sun in the simulation. This figure can be compared to Fig.~1 of \citet{pawlowski15newsat}.}
   \label{fig:aitoff}
\end{figure*}

Figure~\ref{fig:m31sun} shows the projection of the particles belonging to M\,31 as seen from the Sun at the current time. It is colored according to the average line-of-sight velocity with the systemic velocity subtracted. Here north is up and west is right to facilitate the comparison to real images of M\,31, such as \fig{m31sp} or~\ref{fig:streams}. We noted several interesting features here: 1) The tidal {structures} around M\,31, formed exclusively by the material coming from the MW, contain a~pronounced planar sub-structure seen edge-on from the Sun, resembling the GPoA. A~remnant of a~dissolved TDG lies in this plane. 2) The radial extent of this plane, as seen in this projection, is 200-400\,kpc. This corresponds to the radial extent of the GPoA (\fig{m31sp}, see also Fig.~3 by \citealp{kroupacjp} for a~projection where it appears larger). 3) The receding part of the M\,31 disk lies in the northern half of the galaxy. This is a~{trivial} consequence of our simulation being set to approximately reproduce the galaxy inclinations. 4) The approaching part of the planar feature in the simulation lies in the same half of M\,31 as its projected spin. This is also the case for the GPoA (\fig{m31sp}). 5) There are clouds of high velocity particles near the M\,31 disk. They are probably particles on eccentric orbits near their pericenters. The angle between the planar feature and the simulated MW disk was around 40\degr, while the GPoA is perpendicular to the MW disk. The simulation thus shows that a~MW-M\,31 encounter can produce a~planar feature in M\,31 resembling the GPoA in several aspects easily. On the other hand, we will demonstrate in a~next paper using restricted three-body simulations that the morphology of the tidal structures formed by such an encounter depends sensitively on the choice of free parameters.

It is thus possible that a~change of some free parameters would affect the morphology of the tidal structures similarly to viewing the simulated M\,31 from another direction. This is what can be seen in Figs.~\ref{fig:simall} (d), \ref{fig:yzproj} and~\ref{fig:xzproj}. We want to point out the similarity of the tidal {structures} around the simulated M\,31 in \fig{simall} (d) to the tidal features in the real M\,31 (\fig{streams}) which have a~similar size, stream-like morphology and are joined to some of the satellites.

The tidal material formed a~less distinct planar substructure around the MW at the current time. Figure~\ref{fig:mwdosedge} shows a~view where both the flattened feature and the modeled MW disk are seen edge-on. The radial extent of this feature is around 200\,kpc. The radial extent of the VPOS is also around 200\,kpc, see Fig.~3 by \citet{pawlowski15newsat}. Unlike the VPOS, this SP candidate is not polar. There are no polar structures in our simulated MW. The angle between the SP candidate around the MW and the MW-M\,31 connecting line was around 10\degr in our simulation, i.e. it was oriented almost edge-on towards M\,31. Such an alignment is even better than that for the real objects. The SP candidate here rotates in the opposite sense than the SP in the simulated M\,31. In the real case, the VPOS and the GPoA rotate in the same sense.


Figure~\ref{fig:aitoff} shows the Aitoff projection of our simulation from the position of the Sun in our Galactic coordinate system at the current time (compare to Fig.~1 of \citealp{pawlowski15newsat}). The particles belonging to M\,31 are shown in green. The red particles belong to MW and have a~vertical distance from its midplane higher than 50\,kpc. They belong mostly to our SP candidate (compare to \fig{mwdosedge}). The remaining MW particles are blue. We also plotted the velocities of the red particles projected on the plane of the sky using the assumed Sun's orbital velocity. One can see from \fig{aitoff} that most of the red particles form a~coherent structures in phase space (i.e., nearby particles have similar velocities). In this regard, our SP candidate was similar to the VPOS. 

{The mass transferred in the simulation to M\,31 was $2\times 10^{9}$\,M$_\sun$. This matches by an order of magnitude the mass of the real M\,31 streams, $8\times 10^{9}$\,M$_\sun$, or the total mass of its halo, $11\times 10^{9}$\,M$_\sun$ \citep{ibata14b}. This is a~surprisingly good match given that the simulation was not tuned for this.} The structures around the simulated MW could be morphologically classified as streams, several of which appear around the real MW (Fig.~5 by \citealp{pawlowski12} shows several examples and their sizes). To obtain the mass of the MW halo in the simulation, we counted the particles further than 50\,kpc off the MW disk midplane. This limit was chosen as the height of what visually appeared as the warped disk, see \fig{mwdosedge}. We found that such particles constitute only 0.086\% of the total MW mass. {The real baryonic halo mass fraction of the MW is around ten times higher, around 1\% \citep{bland16}. Maybe another choice of the free parameters or a~less simplified simulation would lead to a~better match. This result can also mean that the MW halo was formed by a~mechanism unrelated to the MW-M\,31 encounter. That the galaxy halo formation mechanism is not universal is suggested by the observational finding  by  \citet{merritt16} that galaxies with a~MW luminosity have a~wide variety of halo mass fractions.}

\begin{figure}
 \resizebox{\hsize}{!}{\includegraphics{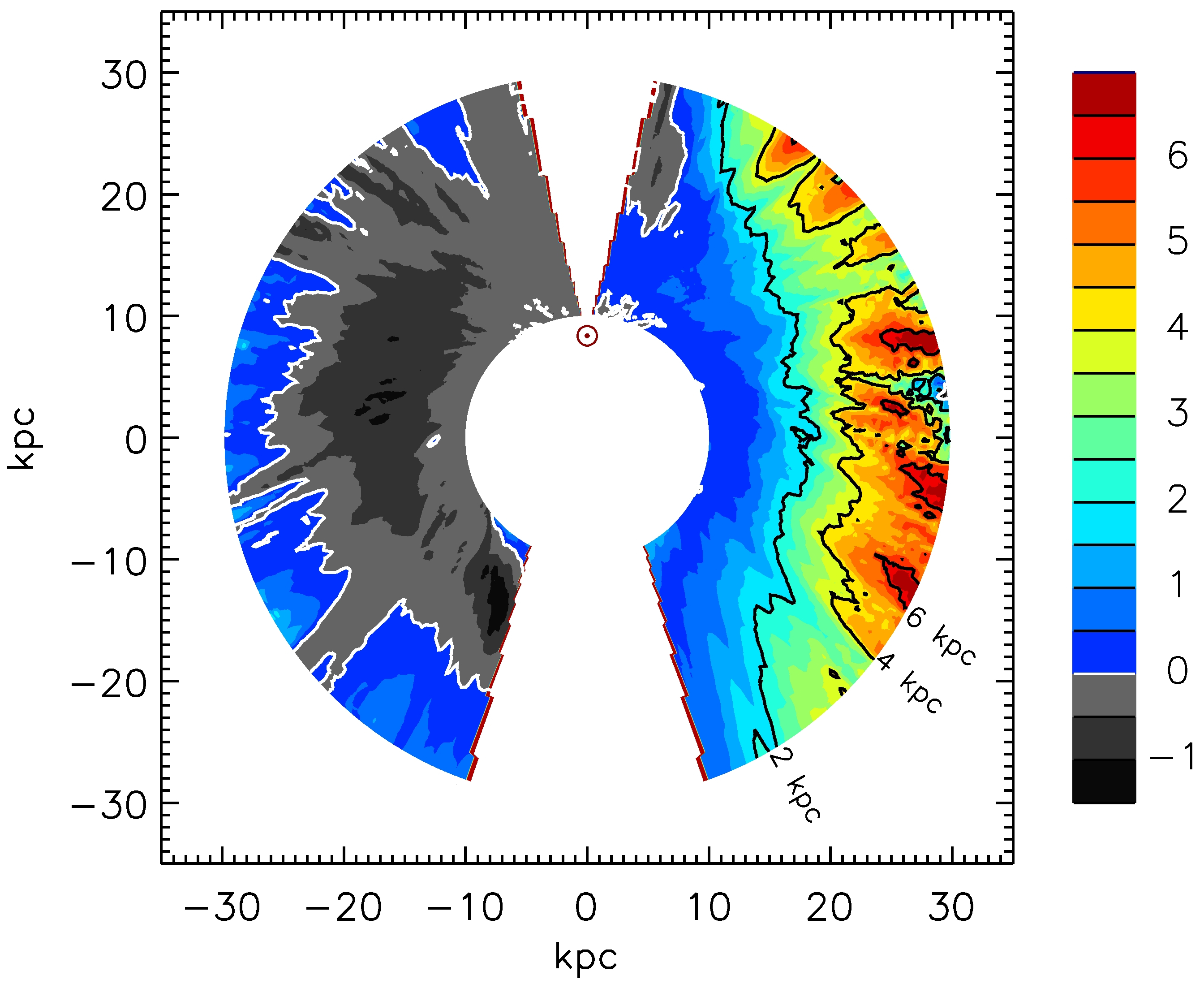}}
 \caption{Observed warp in the outer MW \ion{H}{I} disk. The color indicates the elevation of the disk above the MW midplane ($b = 0\degr$). The $\sun$ symbol marks the position of the Sun. Image courtesy Leo Blitz (adapted Fig.~2 from \citealp{levine06}).}
 \label{fig:warpobs}
\end{figure}

\begin{figure}
 \resizebox{\hsize}{!}{\includegraphics{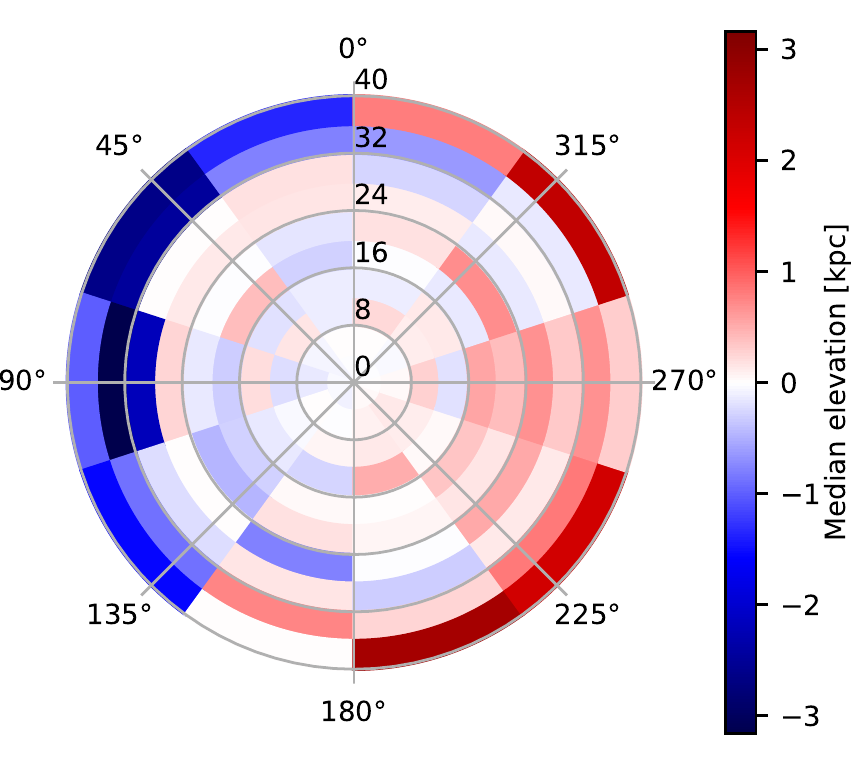}}
 \caption{Vertical elevation of the MW particles above the plane fitted to the inner 15\,kpc of the galaxy. The Sun lies at 0\degr and 8.5\,kpc.{ The radial scale in kiloparsecs is indicated on the vertical axis.} The simulated MW disk is warped similarly to the disk of the real MW, see \fig{warpobs}.} 
 \label{fig:warp}
\end{figure}

\begin{figure}
 \resizebox{\hsize}{!}{\includegraphics{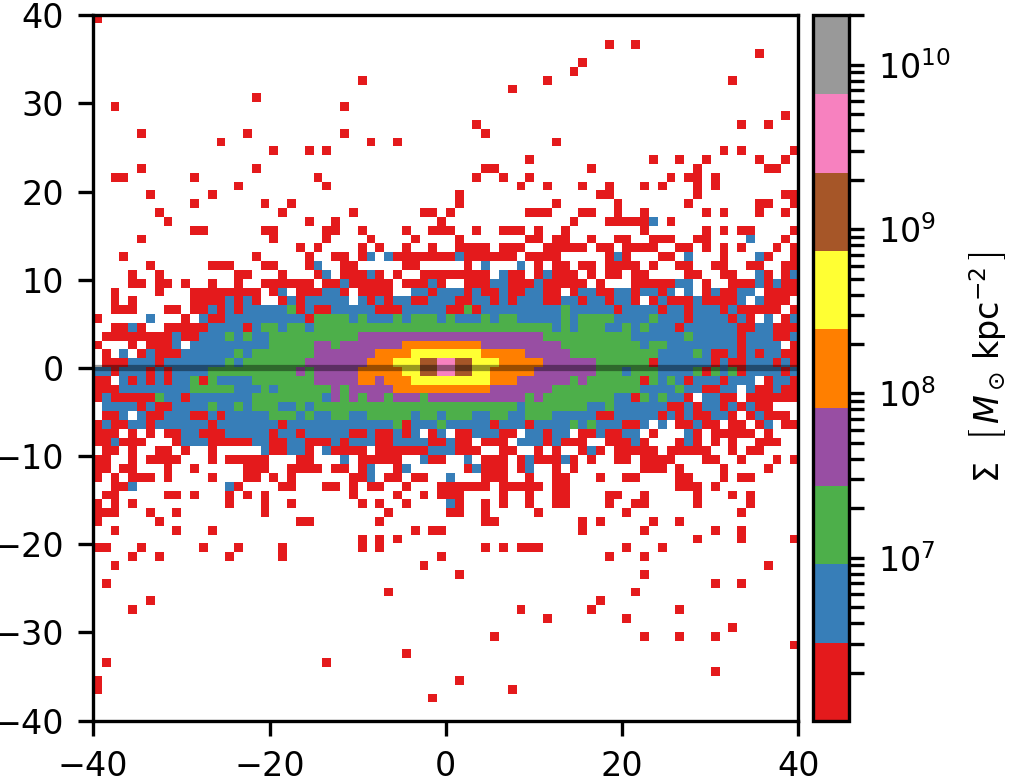}}
 \caption{{Edge-on view of the warp in the simulated MW. The horizontal line marks the MW midplane.}} 
 \label{fig:warpeo}
\end{figure}

The outer regions of our simulated MW were warped as shown in \fig{warp}. This figure shows the median elevation of particles above the {MW midplane} calculated using the bins indicated in the figure by the colored annular sectors. The MW is seen from the north Galactic pole here. The Sun lies at 0\degr and 8.5\,kpc.{ An edge-on view maximizing the visibility of the warp is shown in \fig{warpeo}.} The disk of the real MW is also warped, see \fig{warpobs} showing the elevation of its \ion{H}{I} disk. The Sun position is marked by its symbol. The elevation here is measured above the agreed Galactic midplane having the Galactic latitude of $b = 0\degr$. We could note several similarities to our simulation here. 1) The Sun lies approximately on the line dividing the raised and lowered halves of the disk and these halves lie on the correct sides of the galaxy. We remind that the position of the Sun in the simulation was defined using the position of M\,31. 2) The range of elevation in the simulation, about -3 to 2\,kpc, was surprisingly close to the real range which is about -1 to 6\,kpc. Note that this elevation was reached in the simulation at somewhat larger radii. {In the context of MOND, the MW warp has already been suggested to originate from the EFE exerted by the Large Magellanic Cloud which can explain both its orientation an approximate magnitude \citep{brada00}. The EFE and the encounter are thus probably shaping the warp together in the context of MOND.}

\begin{figure}
 \resizebox{\hsize}{!}{\includegraphics{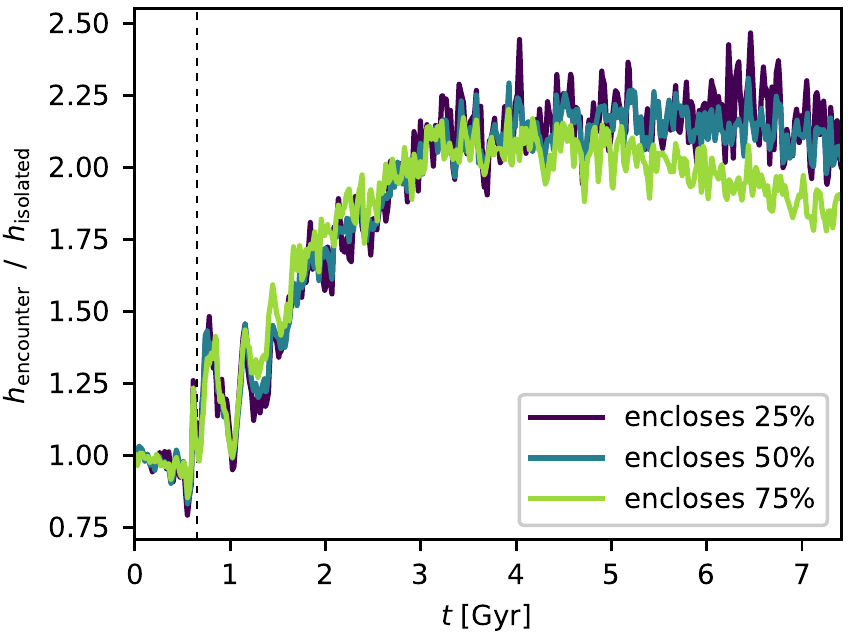}}
 \caption{{The growth of the thickness of the MW with time in the simulation of the encounter. The thickness is displayed normed to the thickness of the MW in the simulation in isolation. The height of the disk was defined as enclosing either 25, 50 or 75\% of the particles in the vertical direction. Only the particles close to the Solar radius (8.5\,kpc) were considered in the calculation. The dashed line marks the instant of the closest MW-M\,31 approach. } 
 \label{fig:heightmw}}
\end{figure}

\begin{figure}
 \resizebox{\hsize}{!}{\includegraphics{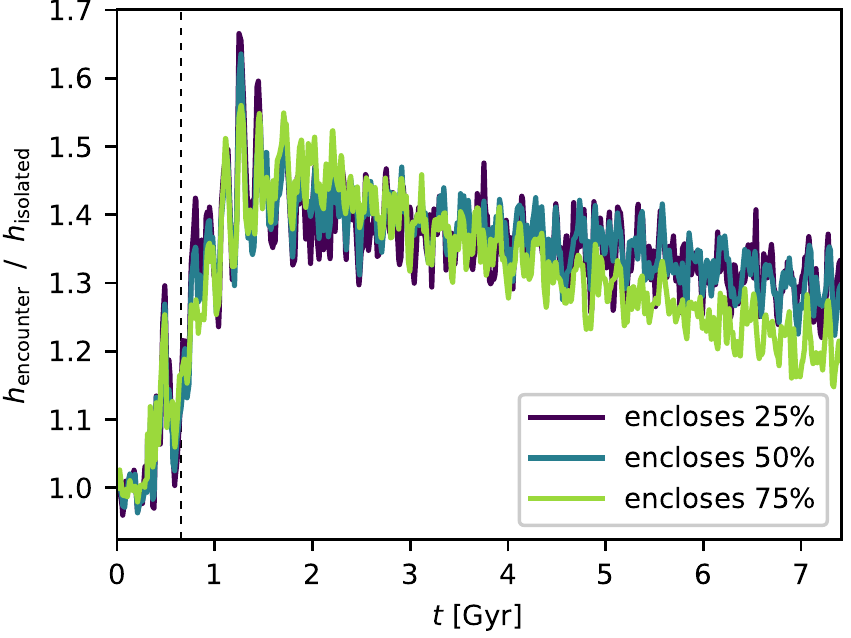}}
 \caption{{The growth of the thickness of M\,31 with time in the simulation of the encounter. The thickness is displayed normed to the thickness of M\,31 in the simulation in isolation. The height of the disk was defined as enclosing either 25, 50 or 75\% of the particles in the vertical direction. Only the particles close to the Solar radius (8.5\,kpc) were considered in the calculation. The dashed line marks the instant of the closest MW-M\,31 approach.} } 
 \label{fig:heightm31}
\end{figure}

{It was suggested by Z13 that the MW-M\,31 encounter could have contributed to the growth of the MW thick disk (but it was probably not the only contribution since thick disks are observed in most disk galaxies, see, e.g.,  \citealp{yoachim06} or \citealp{comeron12}). We thus studied the evolution of the thicknesses of the galaxies in our simulation. We defined the thickness of a~disk as the height of a~layer centered on the galaxy midplane enclosing a~given fraction of particles. We considered only the particles  whose projections to the galactic midplane were near to the Solar radius, namely with a~distance between 7.5 and 9.5\,kpc from the galactic center. Figure~\ref{fig:heightmw} (\ref{fig:heightm31}) shows the evolution of the MW (M\,31) thickness for the threshold heights enclosing 25, 50 or 75\% of the particles. The thickness was divided for every time by the analogous thickness extracted from a~simulation where the respective galaxy evolved in isolation in order to filter-out the secular thickness growth. Only the time period between the simulation start and the current time is displayed. The vertical dashed line marks the time of the closest approach. A~galaxy encounter thus adds on the list mechanisms that can cause a~growth of the disk thickness (a review of thick disk formation mechanisms is given in \citealp{minchev12}; see also \citealp{kroupa02} for a~mechanism which may play a~role when the star-formation rate of a~thin disk is elevated, e.g. through an encounter). Going into details of this mechanism is beyond the scope of this paper.}



%
%
%
%
%

\section{Discussion} \label{sec:disc}

\subsection{Simplifications in the simulation} \label{sec:simpl}
Let us discuss some simplifications of our simulation and estimate their consequences.

How would the large-scale external field affect the encounter? Let us estimate the radii where the external acceleration is comparable to the acceleration from our galaxies. For a~point mass $M$, the deep-MOND limit acceleration is $a_\mathrm{M} = \sqrt{GMa_0}/r$ \citep{milg83a}. This is equal to the external acceleration $a_\mathrm{e}$ at the isolation radius
\begin{equation}
r_\mathrm{isol} = \sqrt{GMa_0}/a_\mathrm{e}.
\label{eq:risol}
\end{equation}
Within this radius, the dynamics of a~galaxy is relatively unaffected by the external field \citep{bm84, milgrom13, milgmondlaws}. The external field acting on the LG is likely a~few hundredths of $a_0$ \citep{famaey07, wu08} and comes mostly from the Virgo and Coma galaxy clusters and from the Great Attractor. The value of $r_\mathrm{isol}$ could, of course, vary with cosmic time -- partly because of the growth of structure and partly because of the possible variations of $a_0$ \citep{milg83a, milga0var, milg17}. For the value $a_\mathrm{e} = 0.03\,a_0$, used by Z13, and our masses of the MW and M\,31 (Eqs.~\ref{eq:mmw} and \ref{eq:mm31}), \equ{risol} gives the MW and M\,31 isolation radii of 280 and 460\,kpc, respectively.  These values are comparable with the sizes of the tidal {structures} in our simulation so that the large-scale gravitational field could affect, for example, the formation of TDGs. { According to Tab.~1 by Z13, including a~realistic external field shifts the pericentric passage into the past by 2-4\,Gyr since the EFE reduces the gravitational attraction between the galaxies. }A~qualitatively same effect has the inclusion of the cosmic expansion. The change of the pericentric velocity and distance would influence the formation of tidal tails. Note that, in order to study whether high-velocity galaxies of the LG could be
reproduced \citep{banik17a, banik17b}, the external field should not be neglected.

A few TDGs formed in our simulation. \citet{wetzstein07} found that TDGs formed in their Newtonian gasless N-body simulations only if the number of particles was too low as a~consequence of particle noise. When gas was included in their simulations, TDGs formed even with a~high number of particles. While the situation can be different in MOND, we recommend considering the TDGs in our simulation{ with reservation} for the moment. Nevertheless, probably all simulations of galaxy encounters in MOND with gas published so far produced several TDGs \citep{tiret07, renaud16, thies16}. There was likely enough gas in the MW and M\,31 around 10\,Gyr ago. For example, \citet{tacconi10} found that the galaxies at $z = 1.2${ with the mass of around $10^{11}$\,M$_\sun$} contain 34\% of their baryonic mass in molecular gas on average. 

Further dwarf galaxies could be descendants of the large gas clouds with masses up to $10^9$\,M$_\sun$ that are observed in galaxies at redshifts greater than about one (e.g., \citealp{zavadsky17, soto17, fisher17}). If some of these clouds were ejected into the tidal tails without being destroyed, they could evolve to dwarf galaxies.


%

We used for the models of our galaxies approximately their current masses, disk sizes, inclinations and density distributions. These quantities were likely different at the time when the pericentric passage is supposed to have happened. There is evidence that galaxies gain mass by accretion of intergalactic gas (e.g., \citealp{sancisi08}). Assuming that the MW and M\,31 evolved along the main star forming sequence, one can estimate their masses in the past. Then it follows from Fig.~3 of \citet{leitner12}, that the MW and M\,31 masses were lower by 30\% than today 7\,Gyr ago, and that these masses were lower by more than 80\% than today before 10\,Gyr. These numbers apply if the \lcdm relation between the redshift and the look-back time works well. It is questionable how this acquired mass and the momentum it brought influenced the trajectories of the galaxies and the inclinations of the disks. Effective radii of galactic disks seem to have evolved only little since the redshift of $z = 1$, while the disk outskirts, as measured by the Petrosian radius, seem to expand substantially{ with the cosmic time} \citep{vandokkum13, sachdeva15}. {If the galaxies had a~greater extent, we expect the tidal arms to be more massive and to be forming more TDGs.} 

It is, of course, possible that several galaxy interactions{, or other mechanisms,} formed the observed SPs in the LG. For example, some of them could have been formed by the mechanism suggested by \citet{hammer13} and \citet{fouquet12}.

\subsection{Comparison to observations}\label{sec:comp}
We saw in \sect{res} that the simulation reproduces some of the features observed in the LG without being tuned for this. To repeat, they are: The encounter led to formation of tidal {structures} around the MW and M\,31 resembling the VPOS and GPoA by sizes and by forming continuous structures in phase space. The tidal structure around the simulated M\,31 contained a~distinct planar substructure. This substructure is similar to the GPoA by pointing towards the MW by its edge, by size, and by the sense of rotation.{ The tidal structures at the simulated M\,31 could be alternatively matched with the tidal features at the real M\,31 because they have a~similar extent and stream-like morphology, and, to within an order of magnitude, also the mass.} The simulated MW disk was warped just like the real one. The nodal line had a~similar orientation with respect to M\,31 and the amplitude of the warp was comparable. {The encounter caused a~vertical thickening of the galactic disks in the simulation. This might have contributed to the thick disk formation, as suggested by Z13.}



{One of the driving questions for our work was whether the observed  SPs in the LG and their special properties are a~simple consequence of the MW-M\,31 encounter in MOND. The VPOS and GPoA have four special properties: 1) The VPOS is perpendicular to the MW disk, 2) The GPoA points by edge towards the MW, 3) The GPoA is perpendicular to the MW disk, 4) The VPOS and GPoA rotate in the same sense. From these properties, our  simulation reproduced only the property 2). We do not know how to modify the simulation to guarantee that the remaining points are reproduced. The SP candidate around our simulated MW is also much less distinct than the VPOS. The two non-satellite dwarf galaxy planes discovered by \citet{pawlowski13} were not reproduced at all. Our simulation thus demonstrates that the MW-M\,31 encounter in MOND does not lead to the formation of all observed properties generically. Future attempts to reproduce them  by the MW-M\,31 encounter in MOND will have to tune the free parameters.}

{Here we summarize the other observations consistent with the past MW-M\,31 encounter that do not relate directly to our simulation.} Deep optical imaging of nearby galaxies with stellar masses similar to that of the MW by \citet{merritt16} revealed that the MW and M\,31 stellar halos are unusually massive and that the halo of M\,31 is exceptionally structured. {It was calculated by Z13 that when the MW-M\,31 encounter was supposed to have happened, the Large and Small Magellanic Clouds were almost in the pericenter of their orbit with respect to the MW.} The classical bulge possessed by M\,31 could be another encounter sign. In \lcdm, classical bulges are supposed to mostly be merger products or to form from giant gas clouds transported to the galaxy center by dynamical friction against the dark matter halo \citep{gadotti12}. We suggest that the bulge could form by the inward gas flow produced by gravity torques during galaxy encounters which accompanies the formation of tidal tails \citep{mihos96, bournaud10}. \citet{bernard15} found that while the stellar populations older than 8\,Gyr constitute over 50\% of the stellar mass in the stream-like structures in M\,31, stellar populations of this age constitute only 38\% of the disk regions. This fact is consistent with the encounter hypothesis since star formation in the tidal material would be probably reduced after diluting a~small piece of the mother galaxy to a~large space. The same can be said about the result by \citet{tenjes17} that the spiral structure in M\,31 has an external origin. The star clusters associated with the VPOS (the Young Halo Globular Clusters) are around 9-12\,Gyr old \citep{pawlowski12}, which agrees with the time since the encounter of 7-11\,Gyr estimated by Z13. A~past encounter between the MW and another galaxy, such as M\,31, can naturally explain the satellites orbiting in the VPOS, but in the opposite sense than the majority of the VPOS members \citep{pawlowski11}.{ The TDGs likely form from a~mix of gas and the stars of the mother galaxy. The metal-enriched gas can cool, collapsing into a~compact core of new stars. The pre-existing stars cannot dynamically cool and their apocenters stay far away from the core. We thus expect an age and metallicity gradient in TDGs with the younger and more metal rich stars at the center. This is indeed observed (e.g., \citealp{battaglia12, kacharov17, okamoto17}). } 

We know about one observation which may be inconsistent with some of the MW satellites being TDGs. We expect that TDGs cannot capture many globular clusters (GCs) of their mother galaxies for the same reason as they cannot contain much dark matter, i.e. the GCs move so fast in the mother galaxy, in equilibrium with its strong gravitational field, that they would quickly escape from the shallow potential wells of the TDGs. Thus, if the observed satellites are TDGs, then they should not contain many GCs that are older than the encounter which formed the TDGs. According to Z13, the MW-M\,31 encounter in MOND occurred 7-11\,Gyr ago. \citet{mackey03} compiled the age estimates of the GCs of  the Fornax and Sagittarius dwarfs based on stellar population models. In their Tab.~3, all 5 Fornax GCs and 3 of 4 Sagittarius GCs are older that 11\,Gyr and 6 of all these 9 GCs are older than 14\,Gyr. The youngest age estimates of the Fornax GCs that we were able to find were published by \citet{deboer16} who derived that the star formation peaked around 12\,Gyr ago for 4 Fornax GCs and 11\,Gyr for the remaining one. {We can see that the age estimates have a~substantial scatter. Another possibility might be }that the MW and M\,31 had lower masses in the past, which is probable \citep{sancisi08}. The effect of mass growth was not taken into account by Z13. {With lower masses, the galaxies would have the encounter a~longer time ago.}

{ There is, of course, the possibility that some of the observed peculiarities in the LG might have an origin unrelated to the MW-M\,31 encounter, even if MOND holds true. }

{The question remains whether the MOND encounter mechanism is able to produce all the observed satellite planes (\sect{intro}). If it is the case, then one should be able to find for every galaxy with a~satellite plane another galaxy which has encountered closely with it. As \citet{kroupacjp} has pointed out, galaxies that have encountered each other in the early universe might appear seemingly unrelated today. For example, if the galaxies interacted 10\,Gyr ago and their average relative velocity was 100\,km\,s$^{-1}$ (compare to \fig{trv}), then they would be separated by 1.0\,Mpc today.  Another requirement for producing a~satellite plane might be a~suitable orbit and inclination of the encountering galaxies. Whether this is a~serious restriction for forming satellite planes is to be clarified by future work.} 

For discussing the formation of the SPs  beyond  the LG by the encounter mechanism, it might be important that the planar structure formed around the simulated M\,31 was a~transient feature in the current model (as can be seen in the video from the simulation): The tidal arm captured by M\,31 formed streams around M\,31 which were changing their shape with time as their particles moved along their Rosetta orbits.

\subsection{Tidal features in galaxies as encounter remnants} \label{sec:tidal}
In the hierarchical galaxy formation scenario, massive galaxies are assembled by merging of lighter galaxies. The observed tidal features in galaxies are often claimed to support this scenario. Our simulation demonstrated, at least in the MOND framework, that some of these tidal features can be formed by galaxy encounters that do not end by a~merger in a~Hubble time. These tidal features are formed by the material exchanged between the encountering galaxies. This material can likely transform into TDGs, some of which might then be accreted to the original galaxies in minor mergers. Thus, observing tidal features and dwarf galaxies being disrupted cannot be considered as unambiguous evidence for the hierarchical galaxy formation scenario. If MOND holds true, a~substantial fraction of tidal features might have been produced by this mechanism because, as our simulation proved, dynamical friction can be insufficient for merging of closely encountering galaxies and numerous close galaxy pairs are observed (see, e.g., Fig.~4 by \citealp{chou12}). This mechanism seems to be much more favored in MOND than in \lcdm since the effective dynamical friction in the latter would probably make to merge soon any galaxies that were able to exchange their baryons (see \sect{lcdm}). A~possible insufficiency of this consideration is that cosmological simulations in MOND have to be done to see what typical galaxy encounter velocities this framework implies: If the encounter velocity is too high, then no material can gain enough momentum to leave its mother galaxy.

\subsection{MW-M\,31 encounter in \lcdm?}\label{sec:lcdm}
It seems unlikely in \lcdm that the MW and M\,31 could have had an encounter that would have any significant effect on their galactic disk because of the efficient dynamical friction. Many, if not all, simulations reproducing the morphology of observed interacting galaxies in \lcdm lead to a~merger in a~few Gyr after the first pericenter (see, e.g., Fig.~1 of \citealp{identikit}). This fast merging gets confirmed when we use Eq.~6 by \citet{jiang14}, which gives the average merging time of two galaxies with given properties according to \lcdm cosmological simulations.  To get a~rough estimate of the merging time for the hypothesized MW-M\,31 past encounter, one can substitute there a~redshift of 1, the virial masses of both galaxies of $10^{11}$\,M$_\sun$ and their separation of 70\,kpc. This leads to the merging time of only 0.9\,Gyr. To get assured about the applicability of this equation to the LG, we used it to obtain the time to the future MW-M\,31 merger. Substituting the redshift of zero, the virial masses of $10^{12}$\,M$_\sun$ and the current distance of 774\,kpc, the equation gave the merging time of 6.6\,Gyr, which agrees well with the MW-M\,31 merger in 5.9\,Gyr in the simulation by \citet{vandermarel12future}. 

{The fast merging stands in the grounds of the classical timing argument for the LG (originally proposed by \citealp{timingargument}). It states that the MW and M\,31 had to have exactly one close encounter in the past and that is at the Big Bang. Then the galaxies were exposed to the Hubble flow and mutual gravitational attraction. If one assumes that the masses of the galaxies did not change over the time and requires that the current galaxy separation and relative velocity are reproduced, the total mass of the LG can be deduced. This provided one of the first indications of the missing mass problem.}

It is interesting to note at this occasion that the relative trajectory of the MW and M\,31 is surprisingly different in MOND and \lcdm, even if both frameworks have to account for the observed dynamics of the galactic disks. When we integrated analytically the MW-M\,31 backwards supposing the MOND two-body-force formula \equ{tbf}, we got the pericenter 7\,Gyr ago (\sect{res}). When we replaced \equ{tbf} by Newton's law and assumed point masses of $2.1\times10^{12}$\,M$_\sun$ (more than most of the recent estimates on the MW and M\,31 virial masses), we got the pericenter 14\,Gyr ago. This is because the gravitational field is different far from the galaxy centers.

\section{Summary}\label{sec:sum}
The paradigm of MOND proved the ability to predict the dynamics of galaxies from the baryonic matter distribution {(see \citealp{famaey12} for a~review)}. When applied to the Local Group (LG), \citet{zhao13} found, using an analytic calculation, that the Milky Way (MW) and Andromeda (M\,31) galaxies had a~close encounter 7-{11}\,Gyr ago. Such an encounter can potentially explain many observed phenomena in the LG. It was suggested that the encounter could explain why many dwarf satellites in the LG concentrate on several planes (the satellite planes, SPs) which have special orientations with respect to the MW, M\,31 and each other. The satellites lying in the planes around the MW and M\,31 (called the VPOS and the GPoA, respectively) moreover orbit in these planes and most of them  in the same sense. Such a~configuration has not been explained satisfactorily in the standard \lcdm model of cosmology yet.

Here we {explored the history of the LG in MOND by performing} the first-ever self-consistent collisionless simulation of the MW-M\,31 encounter in MOND using the publicly available adaptive-mesh-refinement code Phantom of RAMSES \citep{por}. We set up the initial condition so that the simulation approximately reproduces, at a~certain time, the observed separation of the galaxies, their relative velocity, effective radii, masses and inclinations (\sect{desc}). Table~\ref{tab:comp} shows the deviations of these quantities from the observed values and \tab{sim} the found initial conditions. The galaxies came through pericenter 6.8 Gyr before reproducing the observed state (\sect{res}). At the pericenter, their separation reached 24\,kpc and had a~relative flyby velocity of 636\,km\,s$^{-1}$. The encounter did not get even close to merging after more than 13\,Gyr after the first encounter when our simulation ends. This is because the extensive and massive dark matter halos do not exist in MOND.

The simulated encounter led to the transfer of 3\% of the MW mass to M\,31 along a~tidal tail (\sect{res}, \fig{simall} (c)). The other parts of the tidal tails were captured by the MW. No particles escaped possibly because of our neglect of the external field acting on the Local Group such that the escape speed is infinite. The encounter formed clouds{ of particles} around the simulated MW and M\,31. The mass of the cloud formed {at the simulated M\,31  was close to the real baryonic halo mass  of M\,31.} The baryonic halo mass fraction came out 10 times smaller than observed for the simulated MW. The clouds of particles around the simulated galaxies had a~stream like structure. They could be interpreted as the tidal streams observed around the MW and M\,31 (\fig{streams}) because of their similar extent and morphology (compare \fig{simall} (d) and \fig{xzproj} to \fig{streams}).

At the time when the simulation reproduced the observed MW-M\,31 separation and relative velocity, the matter transferred to the simulated M\,31 was forming a~planar structure resembling the GPoA (Figs.~\ref{fig:m31sun} and~\ref{fig:m31sp}) by size, by being oriented by its edge toward the simulated MW and by the same sense of rotation. A~less distinct flattened structure formed at the simulated MW with the extent of the VPOS (\fig{mwdosedge}). 

The encounter induced a~disk warp in the simulated MW very similar to that observed (Figs.~\ref{fig:warp} and~\ref{fig:warpobs}): The zero-elevation line had the right orientation with respect to the direction toward M\,31 and the warp had the right magnitude. {The encounter induced a~thickening of the galaxies in the simulation. The real encounter thus might have contributed to the formation of the thick disks in the MW and M\,31.}


On the other hand, not all properties of the VPOS and the GPoA were reproduced, for example a~distinct SP around the MW perpendicular to its galactic disk was missing, or the planar feature around the simulated M\,31 was not perpendicular to the MW galactic disk as it is the case with the GPoA. The two planes formed by non-satellite LG dwarfs discovered by \citet{pawlowski13} were not reproduced at all. 

Here we see that the MW-M\,31 encounter in MOND has the potential to explain many peculiarities of the LG{ while the match is not perfect at this point.} Our simulation included various simplifications (\sect{simpl}, e.g., we neglected the cosmic expansion, the mass growth of the galaxies, or the gas in the galaxies). {Future investigations should address whether the simulations of the MW-M\,31 encounter in MOND can be made reproducing the observations precisely.} {A recent work by \citet{banik18} has shown that one can produce planes of rotating tidal debris by the encounter in restricted-three-body simulations with correct orbital poles.} Some peculiarities of the LG can be unrelated to the MW-M\,31 encounter, even if MOND is correct.

Apart from {the} main line of our investigation, the simulation revealed the unexpected possibility that tidal features in galaxies (not only in the LG) can be formed by mass exchange between encountering non-merging galaxies (\sect{tidal}). Tidal features observed in galaxies thus cannot be considered as unambiguous proof that galaxies grow predominantly by merging.

\begin{acknowledgements}
The authors thank Marcel Pawlowski, Leo Blitz, Dougal Mackey and Annette Ferguson for the provided figures. We appreciate the help by Guillaume Thomas with processing the RAMSES output.  {We thank Indranil Banik and Mordehai Milgrom for their useful comments and suggestions.}

\end{acknowledgements}

\bibliographystyle{aa}
\bibliography{citace}

\begin{thebibliography}{131}
\expandafter\ifx\csname natexlab\endcsname\relax\def\natexlab#1{#1}\fi

\bibitem[{{Angus}(2008)}]{angus08dwarf}
{Angus}, G.~W. 2008, \mnras, 387, 1481

\bibitem[{{Angus} {et~al.}(2012){Angus}, {van der Heyden}, {Famaey}, {Gentile},
  {McGaugh}, \& {de Blok}}]{angus12}
{Angus}, G.~W., {van der Heyden}, K.~J., {Famaey}, B., {et~al.} 2012, \mnras,
  421, 2598

\bibitem[{{Banik} \& {Zhao}(2015)}]{banik15}
{Banik}, I. \& {Zhao}, H. 2015, ArXiv e-prints [\eprint[arXiv]{1509.08457}]

\bibitem[{{Banik} \& {Zhao}(2017{\natexlab{a}})}]{banik17a}
{Banik}, I. \& {Zhao}, H. 2017{\natexlab{a}}, ArXiv e-prints
  [\eprint[arXiv]{1701.06559}]

\bibitem[{{Banik} \& {Zhao}(2017{\natexlab{b}})}]{banik17b}
{Banik}, I. \& {Zhao}, H. 2017{\natexlab{b}}, \mnras, 467, 2180

\bibitem[{{Banik} \& {Zhao}(2018)}]{banik18}
{Banik}, I. \& {Zhao}, H. 2018, ArXiv e-prints [\eprint[arXiv]{1802.00440}]

\bibitem[{{Barnes} \& {Hernquist}(1992)}]{barnes92}
{Barnes}, J.~E. \& {Hernquist}, L. 1992, \nat, 360, 715

\bibitem[{{Battaglia} {et~al.}(2012){Battaglia}, {Irwin}, {Tolstoy}, {de Boer},
  \& {Mateo}}]{battaglia12}
{Battaglia}, G., {Irwin}, M., {Tolstoy}, E., {de Boer}, T., \& {Mateo}, M.
  2012, \apjl, 761, L31

\bibitem[{{Begeman} {et~al.}(1991){Begeman}, {Broeils}, \&
  {Sanders}}]{begeman91}
{Begeman}, K.~G., {Broeils}, A.~H., \& {Sanders}, R.~H. 1991, \mnras, 249, 523

\bibitem[{{Bekenstein} \& {Milgrom}(1984)}]{bm84}
{Bekenstein}, J. \& {Milgrom}, M. 1984, \apj, 286, 7

\bibitem[{{Bernard} {et~al.}(2015){Bernard}, {Ferguson}, {Richardson}, {Irwin},
  {Barker}, {Hidalgo}, {Aparicio}, {Chapman}, {Ibata}, {Lewis}, {McConnachie},
  \& {Tanvir}}]{bernard15}
{Bernard}, E.~J., {Ferguson}, A.~M.~N., {Richardson}, J.~C., {et~al.} 2015,
  \mnras, 446, 2789

\bibitem[{{Bland-Hawthorn} \& {Gerhard}(2016)}]{bland16}
{Bland-Hawthorn}, J. \& {Gerhard}, O. 2016, \araa, 54, 529

\bibitem[{{Bournaud}(2010)}]{bournaud10}
{Bournaud}, F. 2010, in Astronomical Society of the Pacific Conference Series,
  Vol. 423, Galaxy Wars: Stellar Populations and Star Formation in Interacting
  Galaxies, ed. B.~{Smith}, J.~{Higdon}, S.~{Higdon}, \& N.~{Bastian}, 177

\bibitem[{{Bovy} \& {Rix}(2013)}]{bovy13}
{Bovy}, J. \& {Rix}, H.-W. 2013, \apj, 779, 115

\bibitem[{{Brada} \& {Milgrom}(2000{\natexlab{a}})}]{satelliteefe}
{Brada}, R. \& {Milgrom}, M. 2000{\natexlab{a}}, \apj, 541, 556

\bibitem[{{Brada} \& {Milgrom}(2000{\natexlab{b}})}]{brada00}
{Brada}, R. \& {Milgrom}, M. 2000{\natexlab{b}}, \apjl, 531, L21

\bibitem[{{Bullock} \& {Boylan-Kolchin}(2017)}]{bullock17}
{Bullock}, J.~S. \& {Boylan-Kolchin}, M. 2017, \araa, 55, 343

\bibitem[{{Candlish}(2016)}]{candlish16}
{Candlish}, G.~N. 2016, \mnras, 460, 2571

\bibitem[{{Candlish} {et~al.}(2015){Candlish}, {Smith}, \&
  {Fellhauer}}]{candlish15}
{Candlish}, G.~N., {Smith}, R., \& {Fellhauer}, M. 2015, \mnras, 446, 1060

\bibitem[{{Carignan} {et~al.}(2006){Carignan}, {Chemin}, {Huchtmeier}, \&
  {Lockman}}]{carignan06}
{Carignan}, C., {Chemin}, L., {Huchtmeier}, W.~K., \& {Lockman}, F.~J. 2006,
  \apjl, 641, L109

\bibitem[{{Carraro}(2015)}]{carraro15}
{Carraro}, G. 2015, Boletin de la Asociacion Argentina de Astronomia La Plata
  Argentina, 57, 138

\bibitem[{{Casas} {et~al.}(2012){Casas}, {Arias}, {Pe{\~n}a Ram{\'{\i}}rez}, \&
  {Kroupa}}]{casas12}
{Casas}, R.~A., {Arias}, V., {Pe{\~n}a Ram{\'{\i}}rez}, K., \& {Kroupa}, P.
  2012, \mnras, 424, 1941

\bibitem[{{Chiboucas} {et~al.}(2013){Chiboucas}, {Jacobs}, {Tully}, \&
  {Karachentsev}}]{chiboucas13}
{Chiboucas}, K., {Jacobs}, B.~A., {Tully}, R.~B., \& {Karachentsev}, I.~D.
  2013, \aj, 146, 126

\bibitem[{{Chou} {et~al.}(2012){Chou}, {Bridge}, \& {Abraham}}]{chou12}
{Chou}, R.~C.~Y., {Bridge}, C.~R., \& {Abraham}, R.~G. 2012, \apj, 760, 113

\bibitem[{{Combes}(2016)}]{combes16}
{Combes}, F. 2016, Galactic Bulges, 418, 413

\bibitem[{{Comer{\'o}n} {et~al.}(2012){Comer{\'o}n}, {Elmegreen}, {Salo},
  {Laurikainen}, {Athanassoula}, {Bosma}, {Knapen}, {Gadotti}, {Sheth}, {Hinz},
  {Regan}, {Gil de Paz}, {Mu{\~n}oz-Mateos}, {Men{\'e}ndez-Delmestre},
  {Seibert}, {Kim}, {Mizusawa}, {Laine}, {Ho}, \& {Holwerda}}]{comeron12}
{Comer{\'o}n}, S., {Elmegreen}, B.~G., {Salo}, H., {et~al.} 2012, \apj, 759, 98

\bibitem[{{Courteau} {et~al.}(2011){Courteau}, {Widrow}, {McDonald},
  {Guhathakurta}, {Gilbert}, {Zhu}, {Beaton}, \& {Majewski}}]{courteau11}
{Courteau}, S., {Widrow}, L.~M., {McDonald}, M., {et~al.} 2011, \apj, 739, 20

\bibitem[{{Crnojevi{\'c}} {et~al.}(2016){Crnojevi{\'c}}, {Sand}, {Spekkens},
  {Caldwell}, {Guhathakurta}, {McLeod}, {Seth}, {Simon}, {Strader}, \&
  {Toloba}}]{crnojevic16}
{Crnojevi{\'c}}, D., {Sand}, D.~J., {Spekkens}, K., {et~al.} 2016, \apj, 823,
  19

\bibitem[{{Dabringhausen} {et~al.}(2016){Dabringhausen}, {Kroupa}, {Famaey}, \&
  {Fellhauer}}]{dabringhausen16}
{Dabringhausen}, J., {Kroupa}, P., {Famaey}, B., \& {Fellhauer}, M. 2016,
  \mnras, 463, 1865

\bibitem[{{de Blok} \& {McGaugh}(1998)}]{deblok98}
{de Blok}, W.~J.~G. \& {McGaugh}, S.~S. 1998, \apj, 508, 132

\bibitem[{{de Boer} \& {Fraser}(2016)}]{deboer16}
{de Boer}, T.~J.~L. \& {Fraser}, M. 2016, \aap, 590, A35

\bibitem[{{Dessauges-Zavadsky} {et~al.}(2017){Dessauges-Zavadsky}, {Schaerer},
  {Cava}, {Mayer}, \& {Tamburello}}]{zavadsky17}
{Dessauges-Zavadsky}, M., {Schaerer}, D., {Cava}, A., {Mayer}, L., \&
  {Tamburello}, V. 2017, \apjl, 836, L22

\bibitem[{{Famaey} \& {Binney}(2005)}]{famaey05}
{Famaey}, B. \& {Binney}, J. 2005, \mnras, 363, 603

\bibitem[{{Famaey} {et~al.}(2007){Famaey}, {Bruneton}, \& {Zhao}}]{famaey07}
{Famaey}, B., {Bruneton}, J.-P., \& {Zhao}, H. 2007, \mnras, 377, L79

\bibitem[{{Famaey} \& {McGaugh}(2012)}]{famaey12}
{Famaey}, B. \& {McGaugh}, S.~S. 2012, Living Reviews in Relativity, 15, 10

\bibitem[{{Ferguson} \& {Mackey}(2016)}]{ferguson16}
{Ferguson}, A.~M.~N. \& {Mackey}, A.~D. 2016, Tidal Streams in the Local Group
  and Beyond, 420, 191

\bibitem[{{Fernando} {et~al.}(2017){Fernando}, {Arias}, {Guglielmo}, {Lewis},
  {Ibata}, \& {Power}}]{fernando17}
{Fernando}, N., {Arias}, V., {Guglielmo}, M., {et~al.} 2017, \mnras, 465, 641

\bibitem[{{Fisher} {et~al.}(2017){Fisher}, {Glazebrook}, {Abraham}, {Damjanov},
  {White}, {Obreschkow}, {Basset}, {Bekiaris}, {Wisnioski}, {Green}, \&
  {Bolatto}}]{fisher17}
{Fisher}, D.~B., {Glazebrook}, K., {Abraham}, R.~G., {et~al.} 2017, \apjl, 839,
  L5

\bibitem[{{Fouquet} {et~al.}(2012){Fouquet}, {Hammer}, {Yang}, {Puech}, \&
  {Flores}}]{fouquet12}
{Fouquet}, S., {Hammer}, F., {Yang}, Y., {Puech}, M., \& {Flores}, H. 2012,
  \mnras, 427, 1769

\bibitem[{{Gadotti}(2012)}]{gadotti12}
{Gadotti}, D.~A. 2012, ArXiv e-prints [\eprint[arXiv]{1208.2295}]

\bibitem[{{Gentile} {et~al.}(2011){Gentile}, {Famaey}, \& {de
  Blok}}]{gentile11}
{Gentile}, G., {Famaey}, B., \& {de Blok}, W.~J.~G. 2011, \aap, 527, A76

\bibitem[{{Hammer} {et~al.}(2013){Hammer}, {Yang}, {Fouquet}, {Pawlowski},
  {Kroupa}, {Puech}, {Flores}, \& {Wang}}]{hammer13}
{Hammer}, F., {Yang}, Y., {Fouquet}, S., {et~al.} 2013, \mnras, 431, 3543

\bibitem[{{Hees} {et~al.}(2016){Hees}, {Famaey}, {Angus}, \&
  {Gentile}}]{hees16}
{Hees}, A., {Famaey}, B., {Angus}, G.~W., \& {Gentile}, G. 2016, \mnras, 455,
  449

\bibitem[{{Ibata} {et~al.}(2014{\natexlab{a}}){Ibata}, {Ibata}, {Famaey}, \&
  {Lewis}}]{ibata14}
{Ibata}, N.~G., {Ibata}, R.~A., {Famaey}, B., \& {Lewis}, G.~F.
  2014{\natexlab{a}}, \nat, 511, 563

\bibitem[{{Ibata} {et~al.}(2015){Ibata}, {Famaey}, {Lewis}, {Ibata}, \&
  {Martin}}]{ibata15}
{Ibata}, R.~A., {Famaey}, B., {Lewis}, G.~F., {Ibata}, N.~G., \& {Martin}, N.
  2015, \apj, 805, 67

\bibitem[{{Ibata} {et~al.}(2013){Ibata}, {Lewis}, {Conn}, {Irwin},
  {McConnachie}, {Chapman}, {Collins}, {Fardal}, {Ferguson}, {Ibata}, {Mackey},
  {Martin}, {Navarro}, {Rich}, {Valls-Gabaud}, \& {Widrow}}]{ibata13}
{Ibata}, R.~A., {Lewis}, G.~F., {Conn}, A.~R., {et~al.} 2013, \nat, 493, 62

\bibitem[{{Ibata} {et~al.}(2014{\natexlab{b}}){Ibata}, {Lewis}, {McConnachie},
  {Martin}, {Irwin}, {Ferguson}, {Babul}, {Bernard}, {Chapman}, {Collins},
  {Fardal}, {Mackey}, {Navarro}, {Pe{\~n}arrubia}, {Rich}, {Tanvir}, \&
  {Widrow}}]{ibata14b}
{Ibata}, R.~A., {Lewis}, G.~F., {McConnachie}, A.~W., {et~al.}
  2014{\natexlab{b}}, \apj, 780, 128

\bibitem[{{Iocco} {et~al.}(2015){Iocco}, {Pato}, \& {Bertone}}]{iocco15}
{Iocco}, F., {Pato}, M., \& {Bertone}, G. 2015, \prd, 92, 084046

\bibitem[{{Jiang} {et~al.}(2014){Jiang}, {Jing}, \& {Han}}]{jiang14}
{Jiang}, C.~Y., {Jing}, Y.~P., \& {Han}, J. 2014, \apj, 790, 7

\bibitem[{{Kacharov} {et~al.}(2017){Kacharov}, {Battaglia}, {Rejkuba}, {Cole},
  {Carrera}, {Fraternali}, {Wilkinson}, {Gallart}, {Irwin}, \&
  {Tolstoy}}]{kacharov17}
{Kacharov}, N., {Battaglia}, G., {Rejkuba}, M., {et~al.} 2017, \mnras, 466,
  2006

\bibitem[{{Kahn} \& {Woltjer}(1959)}]{timingargument}
{Kahn}, F.~D. \& {Woltjer}, L. 1959, \apj, 130, 705

\bibitem[{{Kroupa}(1997)}]{kroupa97}
{Kroupa}, P. 1997, \na, 2, 139

\bibitem[{{Kroupa}(2002)}]{kroupa02}
{Kroupa}, P. 2002, \mnras, 330, 707

\bibitem[{{Kroupa}(2012)}]{kroupa12}
{Kroupa}, P. 2012, \pasa, 29, 395

\bibitem[{{Kroupa}(2015)}]{kroupacjp}
{Kroupa}, P. 2015, Canadian Journal of Physics, 93, 169

\bibitem[{{Kroupa} {et~al.}(2005){Kroupa}, {Theis}, \& {Boily}}]{kroupa05}
{Kroupa}, P., {Theis}, C., \& {Boily}, C.~M. 2005, \aap, 431, 517

\bibitem[{{Leitner}(2012)}]{leitner12}
{Leitner}, S.~N. 2012, \apj, 745, 149

\bibitem[{{Lelli} {et~al.}(2017){Lelli}, {McGaugh}, {Schombert}, \&
  {Pawlowski}}]{lelli17}
{Lelli}, F., {McGaugh}, S.~S., {Schombert}, J.~M., \& {Pawlowski}, M.~S. 2017,
  \apj, 836, 152

\bibitem[{{Levine} {et~al.}(2006){Levine}, {Blitz}, \& {Heiles}}]{levine06}
{Levine}, E.~S., {Blitz}, L., \& {Heiles}, C. 2006, \apj, 643, 881

\bibitem[{{Libeskind} {et~al.}(2016){Libeskind}, {Guo}, {Tempel}, \&
  {Ibata}}]{libeskind16}
{Libeskind}, N.~I., {Guo}, Q., {Tempel}, E., \& {Ibata}, R. 2016, \apj, 830,
  121

\bibitem[{{L{\"u}ghausen} {et~al.}(2015){L{\"u}ghausen}, {Famaey}, \&
  {Kroupa}}]{por}
{L{\"u}ghausen}, F., {Famaey}, B., \& {Kroupa}, P. 2015, Canadian Journal of
  Physics, 93, 232

\bibitem[{{L{\"u}ghausen} {et~al.}(2013){L{\"u}ghausen}, {Famaey}, {Kroupa},
  {Angus}, {Combes}, {Gentile}, {Tiret}, \& {Zhao}}]{lughausen13}
{L{\"u}ghausen}, F., {Famaey}, B., {Kroupa}, P., {et~al.} 2013, ArXiv e-prints
  [\eprint[arXiv]{1304.4931}]

\bibitem[{{Lynden-Bell}(1976)}]{lynden-bell76}
{Lynden-Bell}, D. 1976, \mnras, 174, 695

\bibitem[{{Mackey} \& {Gilmore}(2003)}]{mackey03}
{Mackey}, A.~D. \& {Gilmore}, G.~F. 2003, \mnras, 340, 175

\bibitem[{{McGaugh} \& {Milgrom}(2013{\natexlab{a}})}]{anddwarfi}
{McGaugh}, S. \& {Milgrom}, M. 2013{\natexlab{a}}, \apj, 766, 22

\bibitem[{{McGaugh} \& {Milgrom}(2013{\natexlab{b}})}]{anddwarfii}
{McGaugh}, S. \& {Milgrom}, M. 2013{\natexlab{b}}, \apj, 775, 139

\bibitem[{{McGaugh}(2008)}]{mcgaugh08}
{McGaugh}, S.~S. 2008, \apj, 683, 137

\bibitem[{{McGaugh} {et~al.}(2016){McGaugh}, {Lelli}, \&
  {Schombert}}]{mcgaugh16}
{McGaugh}, S.~S., {Lelli}, F., \& {Schombert}, J.~M. 2016, Physical Review
  Letters, 117, 201101

\bibitem[{{McGaugh} \& {Wolf}(2010)}]{mcgaugh10}
{McGaugh}, S.~S. \& {Wolf}, J. 2010, \apj, 722, 248

\bibitem[{{Merritt} {et~al.}(2016){Merritt}, {van Dokkum}, {Abraham}, \&
  {Zhang}}]{merritt16}
{Merritt}, A., {van Dokkum}, P., {Abraham}, R., \& {Zhang}, J. 2016, \apj, 830,
  62

\bibitem[{{Metz} {et~al.}(2007){Metz}, {Kroupa}, \& {Jerjen}}]{metz07}
{Metz}, M., {Kroupa}, P., \& {Jerjen}, H. 2007, \mnras, 374, 1125

\bibitem[{{Metz} {et~al.}(2009{\natexlab{a}}){Metz}, {Kroupa}, \&
  {Jerjen}}]{metz09}
{Metz}, M., {Kroupa}, P., \& {Jerjen}, H. 2009{\natexlab{a}}, \mnras, 394, 2223

\bibitem[{{Metz} {et~al.}(2008){Metz}, {Kroupa}, \& {Libeskind}}]{metz08}
{Metz}, M., {Kroupa}, P., \& {Libeskind}, N.~I. 2008, \apj, 680, 287

\bibitem[{{Metz} {et~al.}(2009{\natexlab{b}}){Metz}, {Kroupa}, {Theis},
  {Hensler}, \& {Jerjen}}]{metz09b}
{Metz}, M., {Kroupa}, P., {Theis}, C., {Hensler}, G., \& {Jerjen}, H.
  2009{\natexlab{b}}, \apj, 697, 269

\bibitem[{{Mihos} \& {Hernquist}(1996)}]{mihos96}
{Mihos}, J.~C. \& {Hernquist}, L. 1996, \apj, 464, 641

\bibitem[{{Milgrom}(1983)}]{milg83a}
{Milgrom}, M. 1983, \apj, 270, 365

\bibitem[{{Milgrom}(1994)}]{milg94c}
{Milgrom}, M. 1994, \apj, 429, 540

\bibitem[{{Milgrom}(2009)}]{milg09}
{Milgrom}, M. 2009, \apj, 698, 1630

\bibitem[{{Milgrom}(2010)}]{qumond}
{Milgrom}, M. 2010, \mnras, 403, 886

\bibitem[{{Milgrom}(2012)}]{milg12}
{Milgrom}, M. 2012, Physical Review Letters, 109, 131101

\bibitem[{{Milgrom}(2013)}]{milgrom13}
{Milgrom}, M. 2013, ArXiv e-prints [\eprint[arXiv]{1305.3516}]

\bibitem[{{Milgrom}(2014)}]{milgmondlaws}
{Milgrom}, M. 2014, \mnras, 437, 2531

\bibitem[{{Milgrom}(2015{\natexlab{a}})}]{milga0var}
{Milgrom}, M. 2015{\natexlab{a}}, \prd, 91, 044009

\bibitem[{{Milgrom}(2015{\natexlab{b}})}]{milgcjp}
{Milgrom}, M. 2015{\natexlab{b}}, Canadian Journal of Physics, 93, 107

\bibitem[{{Milgrom}(2017)}]{milg17}
{Milgrom}, M. 2017, ArXiv e-prints [\eprint[arXiv]{1703.06110}]

\bibitem[{{Milgrom} \& {Sanders}(2003)}]{milg03}
{Milgrom}, M. \& {Sanders}, R.~H. 2003, \apjl, 599, L25

\bibitem[{{Milgrom} \& {Sanders}(2007)}]{milg07}
{Milgrom}, M. \& {Sanders}, R.~H. 2007, \apjl, 658, L17

\bibitem[{{Minchev} {et~al.}(2012){Minchev}, {Famaey}, {Quillen}, {Dehnen},
  {Martig}, \& {Siebert}}]{minchev12}
{Minchev}, I., {Famaey}, B., {Quillen}, A.~C., {et~al.} 2012, \aap, 548, A127

\bibitem[{{M{\"u}ller} {et~al.}(2016){M{\"u}ller}, {Jerjen}, {Pawlowski}, \&
  {Binggeli}}]{muller16}
{M{\"u}ller}, O., {Jerjen}, H., {Pawlowski}, M.~S., \& {Binggeli}, B. 2016,
  \aap, 595, A119

\bibitem[{{M{\"u}ller} {et~al.}(2018){M{\"u}ller}, {Pawlowski}, {Jerjen}, \&
  {Lelli}}]{muller18}
{M{\"u}ller}, O., {Pawlowski}, M.~S., {Jerjen}, H., \& {Lelli}, F. 2018,
  \textit{to appear in Science}, ArXiv e-prints [\eprint[arXiv]{1802.00081}]

\bibitem[{{M{\"u}ller} {et~al.}(2017){M{\"u}ller}, {Scalera}, {Binggeli}, \&
  {Jerjen}}]{muller17}
{M{\"u}ller}, O., {Scalera}, R., {Binggeli}, B., \& {Jerjen}, H. 2017, \aap,
  602, A119

\bibitem[{{Nipoti} {et~al.}(2008){Nipoti}, {Ciotti}, {Binney}, \&
  {Londrillo}}]{nipoti08}
{Nipoti}, C., {Ciotti}, L., {Binney}, J., \& {Londrillo}, P. 2008, \mnras, 386,
  2194

\bibitem[{{Okamoto} {et~al.}(2017){Okamoto}, {Arimoto}, {Tolstoy}, {Jablonka},
  {Irwin}, {Komiyama}, {Yamada}, \& {Onodera}}]{okamoto17}
{Okamoto}, S., {Arimoto}, N., {Tolstoy}, E., {et~al.} 2017, \mnras, 467, 208

\bibitem[{{Pawlowski}(2016)}]{pawlowski16}
{Pawlowski}, M.~S. 2016, \mnras, 456, 448

\bibitem[{{Pawlowski} {et~al.}(2017){Pawlowski}, {Dabringhausen}, {Famaey},
  {Flores}, {Hammer}, {Hensler}, {Ibata}, {Kroupa}, {Lewis}, {Libeskind},
  {McGaugh}, {Merritt}, {Puech}, \& {Yang}}]{pawlowski17}
{Pawlowski}, M.~S., {Dabringhausen}, J., {Famaey}, B., {et~al.} 2017,
  Astronomische Nachrichten, 338, 854

\bibitem[{{Pawlowski} {et~al.}(2015{\natexlab{a}}){Pawlowski}, {Famaey},
  {Merritt}, \& {Kroupa}}]{pawlowski15persist}
{Pawlowski}, M.~S., {Famaey}, B., {Merritt}, D., \& {Kroupa}, P.
  2015{\natexlab{a}}, \apj, 815, 19

\bibitem[{{Pawlowski} \& {Kroupa}(2013)}]{pawlowski13vpos}
{Pawlowski}, M.~S. \& {Kroupa}, P. 2013, \mnras, 435, 2116

\bibitem[{{Pawlowski} {et~al.}(2011){Pawlowski}, {Kroupa}, \& {de
  Boer}}]{pawlowski11}
{Pawlowski}, M.~S., {Kroupa}, P., \& {de Boer}, K.~S. 2011, \aap, 532, A118

\bibitem[{{Pawlowski} {et~al.}(2013){Pawlowski}, {Kroupa}, \&
  {Jerjen}}]{pawlowski13}
{Pawlowski}, M.~S., {Kroupa}, P., \& {Jerjen}, H. 2013, \mnras, 435, 1928

\bibitem[{{Pawlowski} {et~al.}(2015{\natexlab{b}}){Pawlowski}, {McGaugh}, \&
  {Jerjen}}]{pawlowski15newsat}
{Pawlowski}, M.~S., {McGaugh}, S.~S., \& {Jerjen}, H. 2015{\natexlab{b}},
  \mnras, 453, 1047

\bibitem[{{Pawlowski} {et~al.}(2012){Pawlowski}, {Pflamm-Altenburg}, \&
  {Kroupa}}]{pawlowski12}
{Pawlowski}, M.~S., {Pflamm-Altenburg}, J., \& {Kroupa}, P. 2012, \mnras, 423,
  1109

\bibitem[{{Ploeckinger} {et~al.}(2018){Ploeckinger}, {Sharma}, {Schaye},
  {Crain}, {Schaller}, \& {Barber}}]{ploeckinger18}
{Ploeckinger}, S., {Sharma}, K., {Schaye}, J., {et~al.} 2018, \mnras, 474, 580

\bibitem[{{Privon} {et~al.}(2013){Privon}, {Barnes}, {Evans}, {Hibbard}, {Yun},
  {Mazzarella}, {Armus}, \& {Surace}}]{identikit}
{Privon}, G.~C., {Barnes}, J.~E., {Evans}, A.~S., {et~al.} 2013, \apj, 771, 120

\bibitem[{{Renaud} {et~al.}(2016){Renaud}, {Famaey}, \& {Kroupa}}]{renaud16}
{Renaud}, F., {Famaey}, B., \& {Kroupa}, P. 2016, \mnras, 463, 3637

\bibitem[{{Richtler} {et~al.}(2011){Richtler}, {Famaey}, {Gentile}, \&
  {Schuberth}}]{richtler11}
{Richtler}, T., {Famaey}, B., {Gentile}, G., \& {Schuberth}, Y. 2011, \aap,
  531, A100

\bibitem[{{Sachdeva} {et~al.}(2015){Sachdeva}, {Gadotti}, {Saha}, \&
  {Singh}}]{sachdeva15}
{Sachdeva}, S., {Gadotti}, D.~A., {Saha}, K., \& {Singh}, H.~P. 2015, \mnras,
  451, 2

\bibitem[{{Salomon} {et~al.}(2016){Salomon}, {Ibata}, {Famaey}, {Martin}, \&
  {Lewis}}]{salomon16}
{Salomon}, J.-B., {Ibata}, R.~A., {Famaey}, B., {Martin}, N.~F., \& {Lewis},
  G.~F. 2016, \mnras, 456, 4432

\bibitem[{{Samurovi{\'c}}(2014)}]{samur14}
{Samurovi{\'c}}, S. 2014, \aap, 570, A132

\bibitem[{{Sancisi} {et~al.}(2008){Sancisi}, {Fraternali}, {Oosterloo}, \& {van
  der Hulst}}]{sancisi08}
{Sancisi}, R., {Fraternali}, F., {Oosterloo}, T., \& {van der Hulst}, T. 2008,
  \aapr, 15, 189

\bibitem[{{Sanders}(1996)}]{sanders96}
{Sanders}, R.~H. 1996, \apj, 473, 117

\bibitem[{{Sanders} \& {Land}(2008)}]{sanders08}
{Sanders}, R.~H. \& {Land}, D.~D. 2008, \mnras, 389, 701

\bibitem[{{Serra} {et~al.}(2010){Serra}, {Angus}, \& {Diaferio}}]{serra10}
{Serra}, A.~L., {Angus}, G.~W., \& {Diaferio}, A. 2010, \aap, 524, A16

\bibitem[{{Soto} {et~al.}(2017){Soto}, {de Mello}, {Rafelski}, {Gardner},
  {Teplitz}, {Koekemoer}, {Ravindranath}, {Grogin}, {Scarlata}, {Kurczynski},
  \& {Gawiser}}]{soto17}
{Soto}, E., {de Mello}, D.~F., {Rafelski}, M., {et~al.} 2017, \apj, 837, 6

\bibitem[{{Tacconi} {et~al.}(2010){Tacconi}, {Genzel}, {Neri}, {Cox}, {Cooper},
  {Shapiro}, {Bolatto}, {Bouch{\'e}}, {Bournaud}, {Burkert}, {Combes},
  {Comerford}, {Davis}, {Schreiber}, {Garcia-Burillo}, {Gracia-Carpio}, {Lutz},
  {Naab}, {Omont}, {Shapley}, {Sternberg}, \& {Weiner}}]{tacconi10}
{Tacconi}, L.~J., {Genzel}, R., {Neri}, R., {et~al.} 2010, \nat, 463, 781

\bibitem[{{Tenjes} {et~al.}(2017){Tenjes}, {Tuvikene}, {Tamm}, {Kipper}, \&
  {Tempel}}]{tenjes17}
{Tenjes}, P., {Tuvikene}, T., {Tamm}, A., {Kipper}, R., \& {Tempel}, E. 2017,
  \aap, 600, A34

\bibitem[{{Thies} {et~al.}(2016){Thies}, {Kroupa}, \& {Famaey}}]{thies16}
{Thies}, I., {Kroupa}, P., \& {Famaey}, B. 2016, ArXiv e-prints
  [\eprint[arXiv]{1606.04942}]

\bibitem[{{Thomas} {et~al.}(2017){Thomas}, {Famaey}, {Ibata}, {L{\"u}ghausen},
  \& {Kroupa}}]{thomas17}
{Thomas}, G.~F., {Famaey}, B., {Ibata}, R., {L{\"u}ghausen}, F., \& {Kroupa},
  P. 2017, \aap, 603, A65

\bibitem[{{Tiret} \& {Combes}(2007)}]{tiret07}
{Tiret}, O. \& {Combes}, F. 2007, \aap, 464, 517

\bibitem[{{Tiret} {et~al.}(2007){Tiret}, {Combes}, {Angus}, {Famaey}, \&
  {Zhao}}]{tiret07b}
{Tiret}, O., {Combes}, F., {Angus}, G.~W., {Famaey}, B., \& {Zhao}, H.~S. 2007,
  \aap, 476, L1

\bibitem[{{Tully} {et~al.}(2015){Tully}, {Libeskind}, {Karachentsev},
  {Karachentseva}, {Rizzi}, \& {Shaya}}]{tully15}
{Tully}, R.~B., {Libeskind}, N.~I., {Karachentsev}, I.~D., {et~al.} 2015,
  \apjl, 802, L25

\bibitem[{{Vakili} {et~al.}(2017){Vakili}, {Kroupa}, \& {Rahvar}}]{vakili17}
{Vakili}, H., {Kroupa}, P., \& {Rahvar}, S. 2017, \apj, 848, 55

\bibitem[{{van der Marel} {et~al.}(2012{\natexlab{a}}){van der Marel}, {Besla},
  {Cox}, {Sohn}, \& {Anderson}}]{vandermarel12future}
{van der Marel}, R.~P., {Besla}, G., {Cox}, T.~J., {Sohn}, S.~T., \&
  {Anderson}, J. 2012{\natexlab{a}}, \apj, 753, 9

\bibitem[{{van der Marel} {et~al.}(2012{\natexlab{b}}){van der Marel},
  {Fardal}, {Besla}, {Beaton}, {Sohn}, {Anderson}, {Brown}, \&
  {Guhathakurta}}]{vandermarel12}
{van der Marel}, R.~P., {Fardal}, M., {Besla}, G., {et~al.} 2012{\natexlab{b}},
  \apj, 753, 8

\bibitem[{{van Dokkum} {et~al.}(2013){van Dokkum}, {Leja}, {Nelson}, {Patel},
  {Skelton}, {Momcheva}, {Brammer}, {Whitaker}, {Lundgren}, {Fumagalli},
  {Conroy}, {F{\"o}rster Schreiber}, {Franx}, {Kriek}, {Labb{\'e}},
  {Marchesini}, {Rix}, {van der Wel}, \& {Wuyts}}]{vandokkum13}
{van Dokkum}, P.~G., {Leja}, J., {Nelson}, E.~J., {et~al.} 2013, \apjl, 771,
  L35

\bibitem[{{Wetzstein} {et~al.}(2007){Wetzstein}, {Naab}, \&
  {Burkert}}]{wetzstein07}
{Wetzstein}, M., {Naab}, T., \& {Burkert}, A. 2007, \mnras, 375, 805

\bibitem[{{Wu} {et~al.}(2008){Wu}, {Famaey}, {Gentile}, {Perets}, \&
  {Zhao}}]{wu08}
{Wu}, X., {Famaey}, B., {Gentile}, G., {Perets}, H., \& {Zhao}, H. 2008,
  \mnras, 386, 2199

\bibitem[{{Yin} {et~al.}(2009){Yin}, {Hou}, {Prantzos}, {Boissier}, {Chang},
  {Shen}, \& {Zhang}}]{yin09}
{Yin}, J., {Hou}, J.~L., {Prantzos}, N., {et~al.} 2009, \aap, 505, 497

\bibitem[{{Yoachim} \& {Dalcanton}(2006)}]{yoachim06}
{Yoachim}, P. \& {Dalcanton}, J.~J. 2006, \aj, 131, 226

\bibitem[{{Zhao} \& {Famaey}(2010)}]{zhao10b}
{Zhao}, H. \& {Famaey}, B. 2010, \prd, 81, 087304

\bibitem[{{Zhao} {et~al.}(2013){Zhao}, {Famaey}, {L{\"u}ghausen}, \&
  {Kroupa}}]{zhao13}
{Zhao}, H., {Famaey}, B., {L{\"u}ghausen}, F., \& {Kroupa}, P. 2013, \aap, 557,
  L3

\bibitem[{{Zhao} {et~al.}(2010){Zhao}, {Li}, \& {Bienaym{\'e}}}]{zhao10}
{Zhao}, H., {Li}, B., \& {Bienaym{\'e}}, O. 2010, \prd, 82, 103001

\end{thebibliography}

\begin{appendix}\label{app:app}
\section{Stability of our models}

\begin{figure}
 \resizebox{\hsize}{!}{\includegraphics{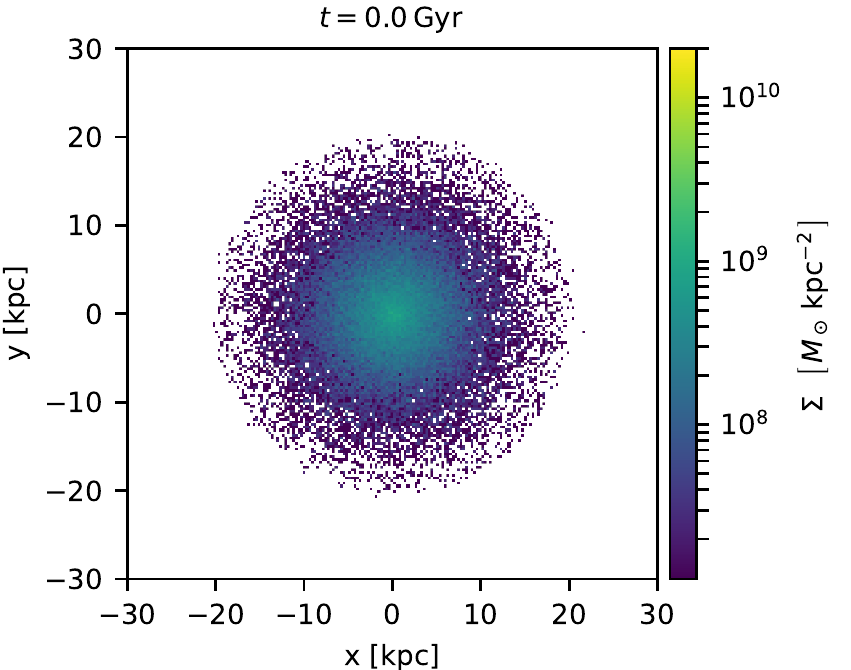}}
 \caption{Milky Way model simulated in isolation at the simulation start (0.0\,Gyr).} 
 \label{fig:mwisolini}
\end{figure}

\begin{figure}
 \resizebox{\hsize}{!}{\includegraphics{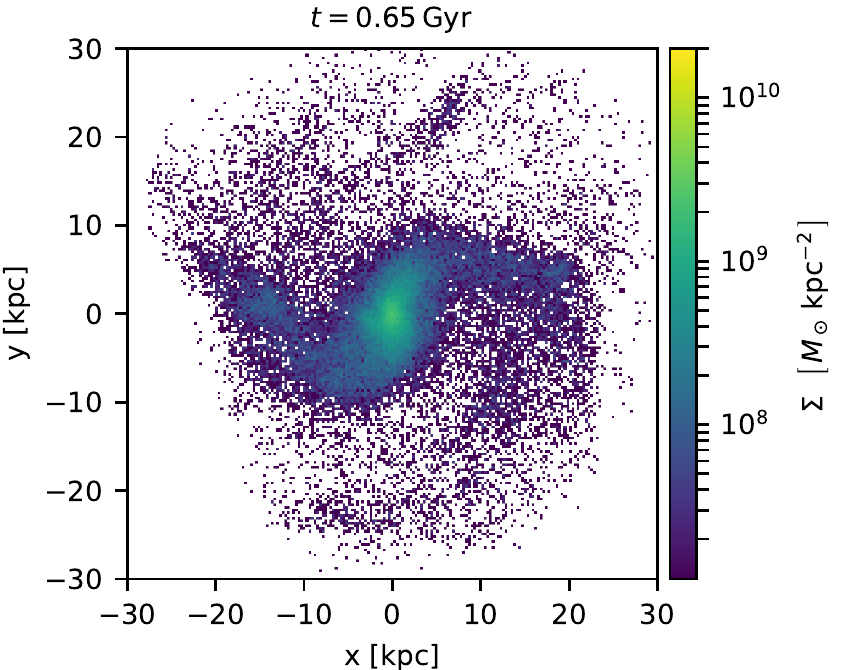}}
 \caption{Milky Way model simulated in isolation at the time when the MW and M\,31 came through the pericenter in the main simulation (0.65\,Gyr).} 
 \label{fig:mwisolperic}
\end{figure}

\begin{figure}
 \resizebox{\hsize}{!}{\includegraphics{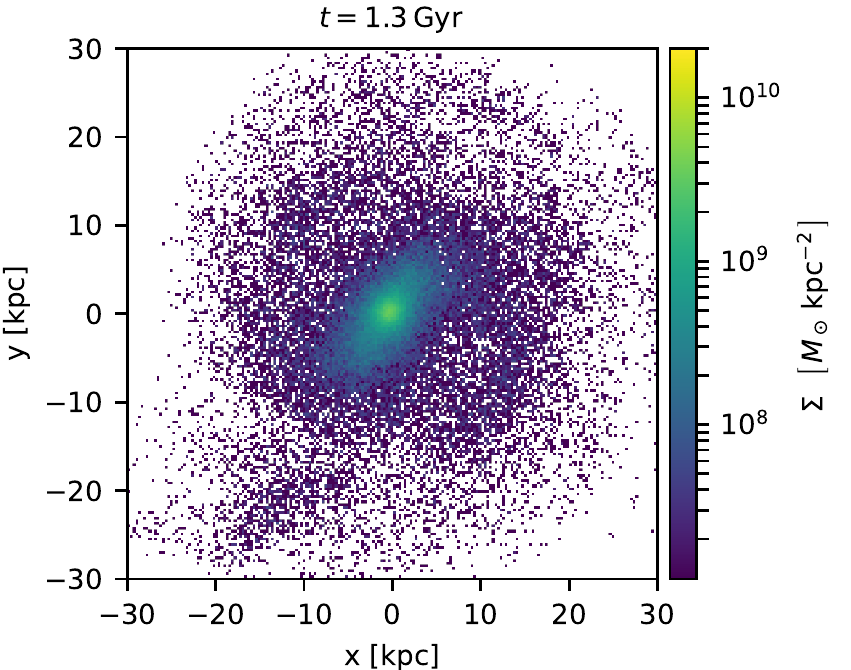}}
 \caption{Milky Way model simulated in isolation at twice the time when the MW and M\,31 came through the pericenter in the main simulation (1.3\,Gyr).} 
 \label{fig:mwisolinterm}
\end{figure}

\begin{figure}
 \resizebox{\hsize}{!}{\includegraphics{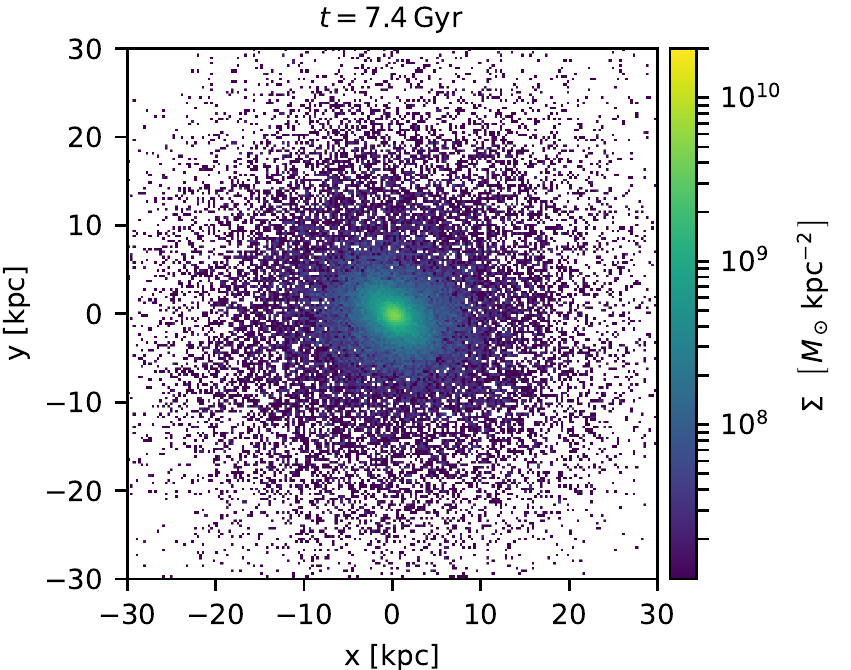}}
 \caption{Milky Way model simulated in isolation at the time corresponding the current time in the main simulation (7.4\,Gyr).} 
 \label{fig:mwisolnow}
\end{figure}
\begin{figure}
 \resizebox{\hsize}{!}{\includegraphics{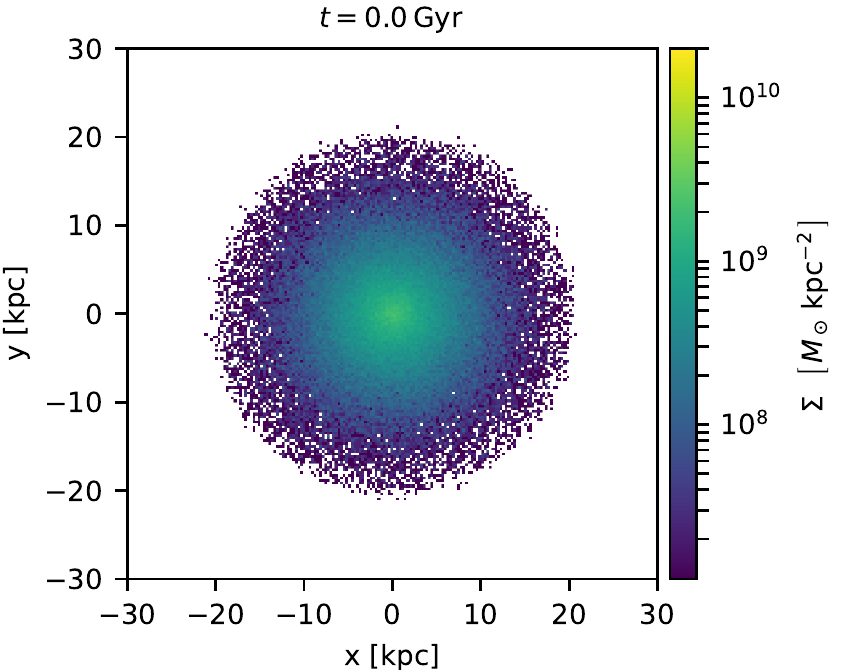}}
 \caption{M\,31 model simulated in isolation at the simulation start (0.0\,Gyr).} 
 \label{fig:m31isolini}
\end{figure}
\begin{figure}
 \resizebox{\hsize}{!}{\includegraphics{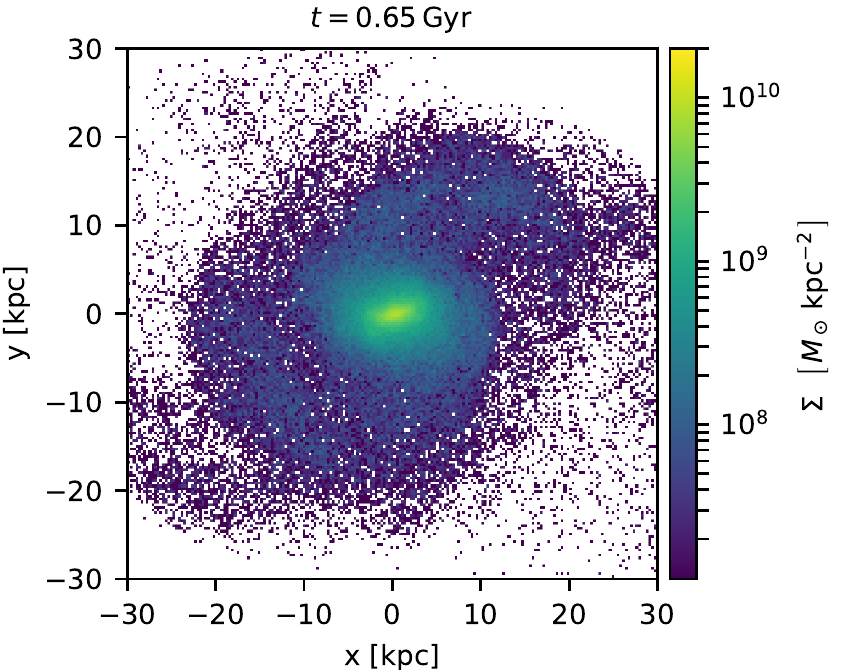}}
 \caption{M\,31 model simulated in isolation at the time when the MW and M\,31 came through the pericenter in the main simulation (0.65\,Gyr).} 
 \label{fig:m31isolperic}
\end{figure}
\begin{figure}
 \resizebox{\hsize}{!}{\includegraphics{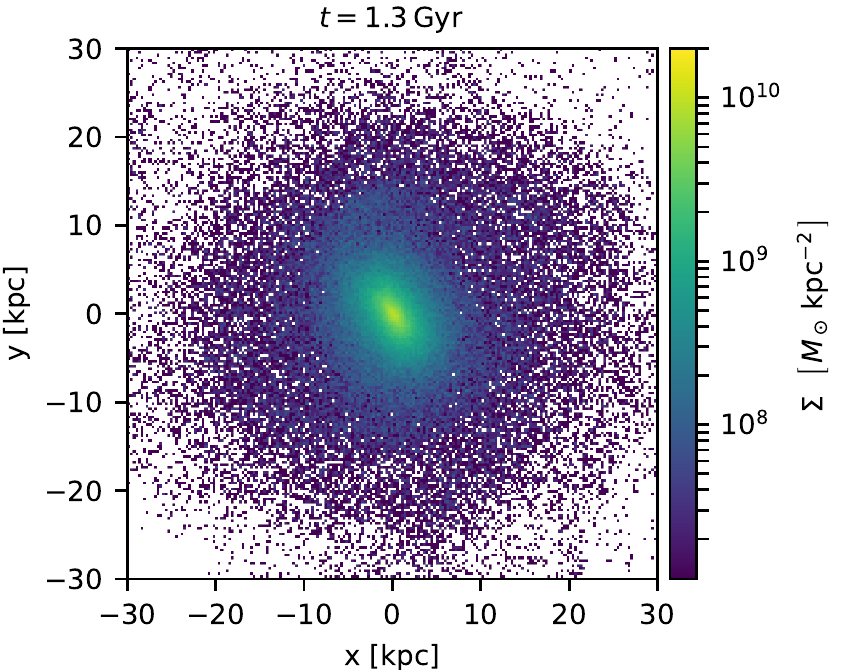}}
 \caption{M\,31 model simulated in isolation at twice the time when the MW and M\,31 came through the pericenter in the main simulation (1.3\,Gyr).} 
 \label{fig:m31isolinterm}
\end{figure}
\begin{figure}
 \resizebox{\hsize}{!}{\includegraphics{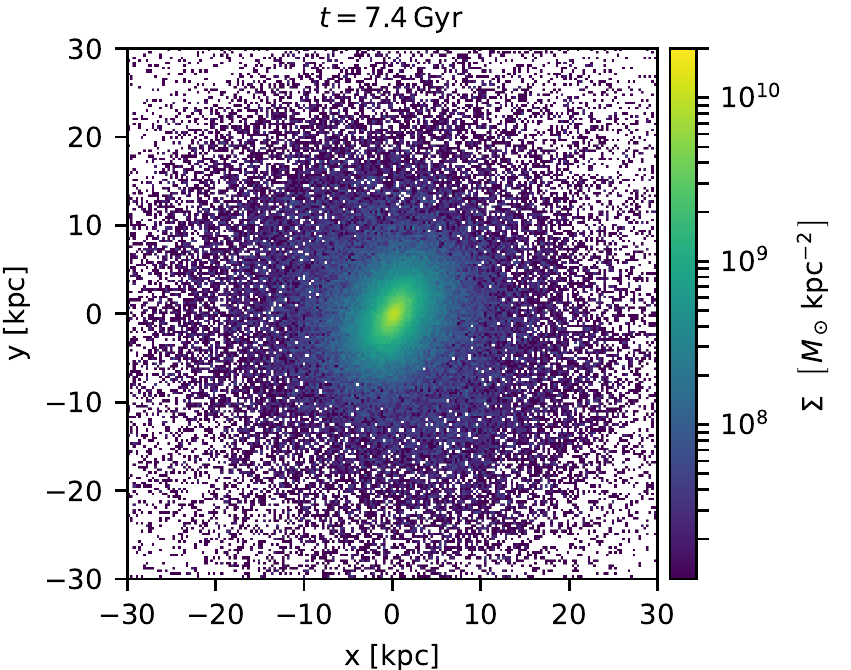}}
 \caption{M\,31 model simulated in isolation at the time corresponding the current time in the main simulation (7.4\,Gyr).} 
 \label{fig:m31isolnow}
\end{figure}
\begin{figure}
 \resizebox{\hsize}{!}{\includegraphics{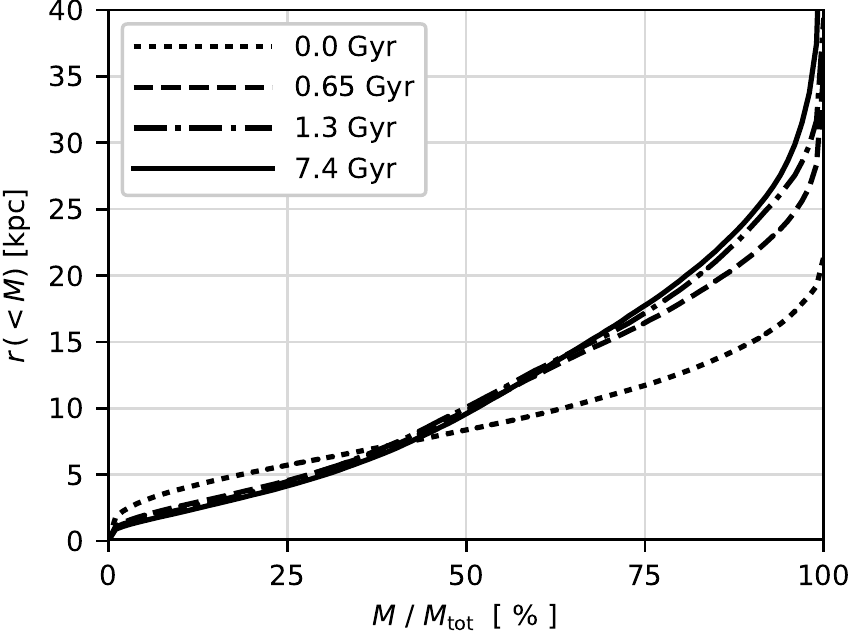}}
 \caption{Lagrangian radius for the model of the MW evolving in isolation as a~function of the enclosed mass. Each line corresponds to the time indicated in the figure legend. } 
 \label{fig:mwisollag}
\end{figure}
\begin{figure}
 \resizebox{\hsize}{!}{\includegraphics{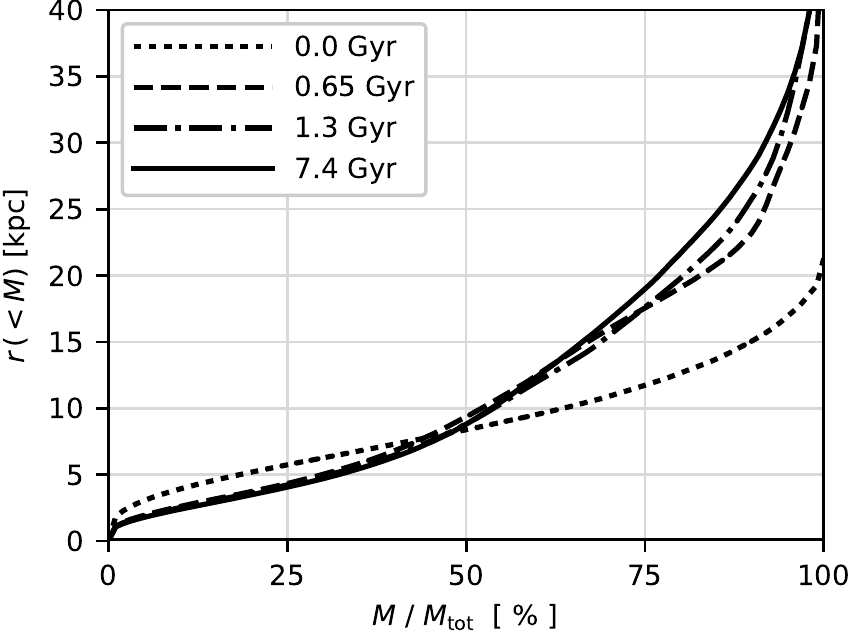}}
 \caption{Lagrangian radius for the model of M\,31 evolving in isolation as a~function of the enclosed mass. Each line corresponds to the time indicated in the figure legend. } 
 \label{fig:m31isollag}
\end{figure}

In order to explore the stability of our initial galaxy models, we let them evolve in isolation using the same computational setup as the main simulation, i.e. that summarized in \tab{sim}. Figures~\ref{fig:mwisolini}--\ref{fig:mwisolnow} show the evolution of the MW model in isolation: \fig{mwisolini} is the start of the simulation (0.0\,Gyr), \fig{mwisolperic} shows the model at the time when the galaxies were in pericenter in the main simulation (0.65\,Gyr) and \fig{mwisolinterm} after twice as much time (1.3\,Gyr). A~stable state was established approximately there, so that the galaxy did not evolve much until the time corresponding to the current time in the main simulation (7.4\,Gyr), see \fig{mwisolnow}. The Figs.~\ref{fig:m31isolini}--\ref{fig:m31isolnow} display the same for M\,31 whose evolution was qualitatively similar. We constructed the plots of the Lagrangian radii at these moments in \fig{mwisollag} for the MW and in \fig{m31isollag} for M\,31.

To summarize, there was a~short period of disk virialization lasting for about 1\,Gyr for both the MW and M\,31 models. During this time, a~bar developed and spiral arms appeared and disappeared. The galaxies evolved only little subsequently. Importantly, there were no escaping particles and the galaxy half-mass radii changed negligibly.

\end{appendix}

\clearpage

\end{document}